\documentclass[revtex4,driverfallback=dvipdfm]{emulateapj}

\usepackage{graphicx}
\usepackage{amsmath}
\usepackage{ulem}
\usepackage{cases}

\usepackage{color}

\usepackage{natbib}

\newcommand{\feoh}{[{\rm Fe} / {\rm H}]}
\newcommand{\msun}{\, M_\odot}
\newcommand{\Zsun}{\, Z_\odot}

\newcommand{\cemps}{CEMP-$s$}

\newcommand{\cfe}{\abra{C}{Fe}}
\newcommand{\coh}{\abra{C}{H}}
\newcommand{\mmd}{M_{\rm md}}
\newcommand{\spr}{{\it s}-process}
\newcommand{\hyp}{\ensuremath{\, \mathchar`- \,}}

\def\abra#1#2{[{\rm #1}/{\rm #2}]}
\def\bfrac#1#2{\left( \frac{#1}{#2} \right)}

\def\bfrac#1#2{\left(\frac{#1}{#2}\right)}

\shorttitle{Faint Supernova}
\shortauthors{Komiya}

\begin{document}
%\nocite{*}
\title{Are Faint Supernovae Responsible for Carbon-Enhanced Metal-Poor Stars?}

\author{Yutaka Komiya\altaffilmark{1}, Takuma Suda\altaffilmark{2,1}, Shimako Yamada\altaffilmark{3}, and Masayuki Y. Fujimoto\altaffilmark{3,4}}
\altaffiltext{1}{Research Center for the Early Universe, University of Tokyo, Hongo 7-3-1, Bunkyo-ku, 113-0033, Tokyo, Japan}
\altaffiltext{2}{Open University of Japan, Wakaba 2-11, Mihama-ku, 261-8586, Chiba, Japan}
\altaffiltext{3}{Graduate School of Science, Hokkaido University, Kita 6 Nishi 10, Kita-ku, Sapporo 060-0810, Japan }
\altaffiltext{4}{Faculty of Engineering, Hokkai-Gakuen University, Asahimachi 4-1-40, Toyohira-ku, Sapporo 060-0810, Japan}

\begin{abstract}
Mixing and fallback models in faint supernova models are supposed to reproduce the abundance patterns of observed carbon-enhanced metal-poor (CEMP) stars in the Galactic halo.
A fine tuning of the model parameters for individual stars is required to reproduce the observed ratios of carbon to iron.
We focus on extremely metal-poor stars formed out of the ejecta from the mixing and fallback models using a chemical evolution model.
Our chemical evolution models take into account the contribution of individual stars to chemical enrichment in host halos together with their evolution in the context of the hierarchical clustering.
Parametrized models of mixing and fallback models for Pop. III faint supernovae are implemented in the chemical evolution models with merger trees to reproduce the observed CEMP stars.
A variety of choices for model parameters on star formation and metal-pollution by faint supernovae is unable to reproduce the observed stars with $\feoh \lesssim -4$ and $\coh \gtrsim -2$, which are the majority of CEMP stars among the lowest metallicity stars.
Only possible solution is to form stars from small ejecta mass, which produces an inconsistent metallicity distribution function. 
We conclude that not all the CEMP stars are explicable by the mixing and fallback models.
We also tested the contribution of binary mass transfers from AGB stars that are also supposed to reproduce the abundances of known CEMP stars.
This model reasonably reproduces the distribution of carbon and iron abundances simultaneously only if we assume that long-period binaries are favored at $\feoh \lesssim -3.5$.
\end{abstract}
 % -- acceleration of particles}

\section{Introduction}

%(EMP)
Extremely metal-poor (EMP) stars in the Milky Way (MW) halo are the survivors of the very early generations of stars. 
They are interesting objects as a key to the understanding of the formation, evolution and explosion of first stars, and also of the early phases of galaxy formation and chemical evolution. 

%(CEMP)
The most prominent feature of EMP stars is a high frequency of carbon-enhanced star which
are classified as carbon-enhanced metal-poor (CEMP) stars \citep[e.g.,][]{beers05}.
The well-accepted criterion of CEMP stars is $\cfe > 0.7$\citep{Aoki07}. 
CEMP stars occupy $20 \hyp 30\%$ of EMP stars \citep[][]{Rossi99, Lee14,Carollo14}. 

\begin{figure}
\includegraphics[width=\columnwidth,pagebox=cropbox]{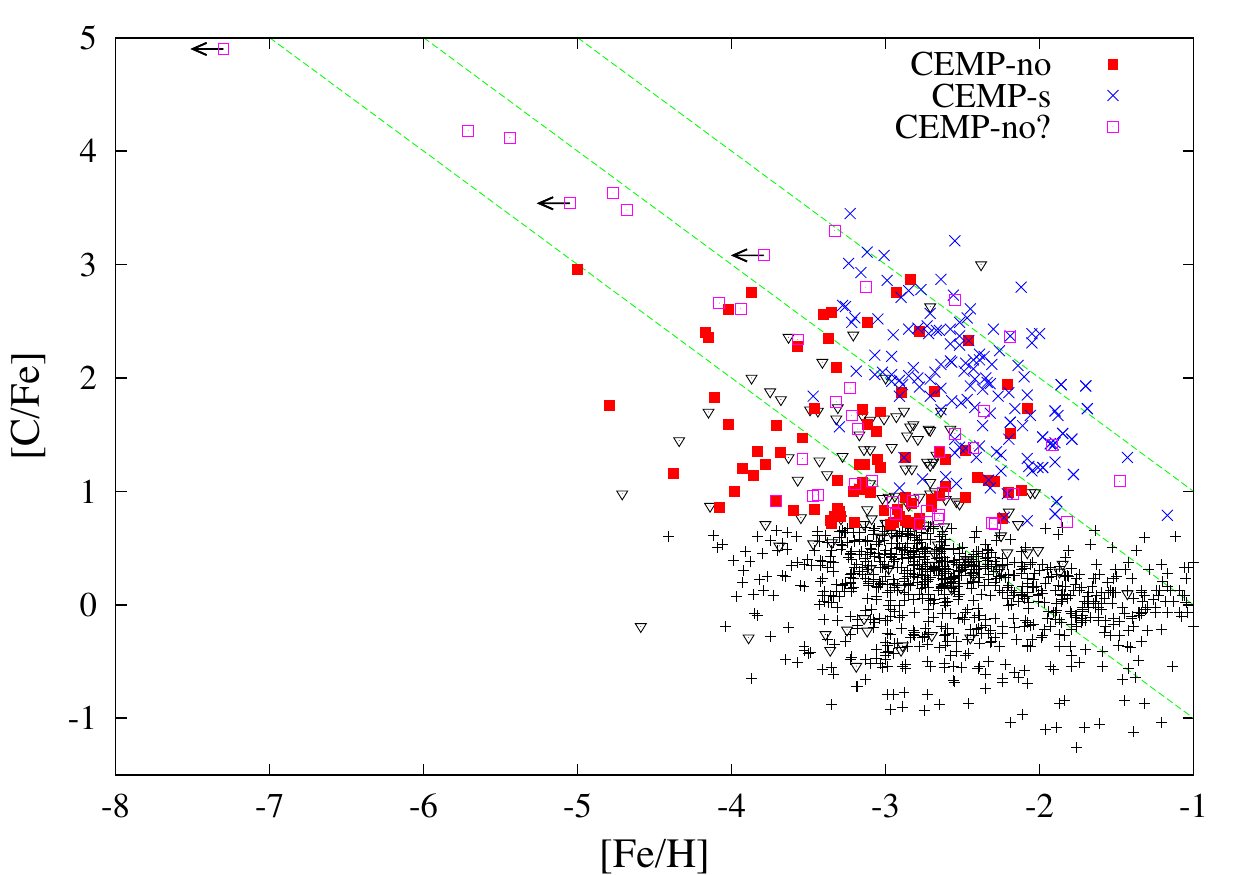}
\caption{Distribution of the metal-poor stars on the metallicity ($\feoh$) and the carbon enhancement ($\cfe$) diagram, where \cemps\ stars (blue crosses), CEMP-no stars (red filled squares), and carbon-normal stars (black plus symbols) are plotted.
Stars with upper limits for carbon abundances are plotted with the inverted triangles.
Magenta open squares denote CEMP stars with upper limits for barium enhancement having $\abra{Ba}{Fe} > 0.5$, or CEMP stars without the measurement of barium abundances. 
%Left pointing triangles
Horizontal arrows stand for upper limits for iron abundances. 
Dashed green diagonal lines are the constant values of $\coh=0, -1, -2$ from top to bottom. 
The observational data are taken from the SAGA database (http://sagadatabase.jp/). 
}\label{obs}
\end{figure}

%(UMP, HMP)
The percentage of carbon-enhanced stars is higher at lower metallicity. 
Among 27 ultra metal-poor (UMP) stars with $\feoh < -4$, registered in the SAGA database which compiles the published observational data of EMP stars \citep{Suda08, Suda11,Yamada13,Suda17a}, 17 stars have carbon abundance with $\cfe > 0.7$ in the April 11, 2018 version.  
Furthermore, in the case of five hyper metal poor (HMP) stars with $\feoh < -5$, all of them show large carbon enhancement.  
All the HMP stars and most of UMP stars have carbon abundance of $-1 \gtrsim \coh \gtrsim -2.5$, as shown in Figure~\ref{obs}. 
We refer to these stars together as carbon-rich UMP (CRUMP) stars in this paper. 

%(CEMP-s/no)
It is known that the majority of CEMP stars also show the enhancement of \spr\ elements \citep[e.g.][]{Aoki07}. 
These stars are called \cemps\ stars (the definition is $\abra{Ba}{Fe} > 0.5$ in this paper), while stars without the enhancement of $s$-process elements are referred to as CEMP-no stars. 

%(Yoon Group I,II,III)
\citet{Yoon16} proposed another classification of CEMP stars according to the carbon and iron abundances. 
They divided CEMP stars into three groups. 
The Group I stars are those with large carbon abundances ($\coh \gtrsim -1.3$) in the metallicity range of $\feoh \simeq -4 \hyp -2$.  
Vast majority of the Group I stars are \cemps\ stars but a dozen CEMP-no stars are also classified into the Group I.  
The Group II population consists of stars with small C enhancement ($\cfe \lesssim +1$) at extremely low metallicity ($\feoh \simeq -5.0 \hyp  -2.5$). 
They are all members of CEMP-no stars.  
Stars with significant C enhancement with $\coh \simeq -1 \hyp -2$ and in the lowest metallicity range ($\feoh \lesssim -3.5$) are classified into the Group III. 
They are almost overlapped with CRUMP stars and are classified as CEMP-no stars, although most of them have measured Ba abundances with only loose upper limits at $\abra{Ba}{Fe} > 0.5$.

%(CEMP-s:binary scenario)
It is now widely accepted that \cemps\ stars are formed through binary mass transfer from intermediate mass companion stars. 
Intermediate mass stars produce carbon and \spr\ elements, and dredge them up in the surface during the TP-AGB phase, though the enhancement of \spr\ elements depends on their progenitor mass \citep[e.g.,][]{Yamada19a}. 
After mass transfer events via stellar wind or the Roche-lobe overflow, secondary companion becomes carbon and \spr\ enhanced stars.  
These secondary stars should be observed as CEMP stars with extinct white dwarf companions. 
Observationally, radial velocity variations were detected for most of the \cemps\ stars that were subject to the monitoring of their radial velocities \citep{Lucatello05,Starkenburg14}. 

%(CEMP-no)
On the other hand, the comprehensive scenario for the origin of CEMP stars including CEMP-no and CRUMP stars is not yet established. 
So far, three major scenarios have been proposed, all of which tried to explain the origin of CRUMP stars (in particular, HE0107-5240), while they are applicable to other CEMP-no stars. 

%(Faint SN scenario)
The first scenario considers peculiar supernovae (SNe) which eject high $\cfe$ material. 
If a star explodes as a core-collapse SN (CCSN), its ejecta will have large $\cfe$ supposing that iron rich material at their inner region falls back on to the central compact object while carbon rich material at the outer layer is blown off. 
If the first supernova was a ``faint SN'' and contains a small amount of iron in the yields, the second generation stars, formed out of these carbon-enhanced ejecta, will be CEMP-no or CRUMP stars. 
  
The `mixing and fallback' model associated with faint SNe was proposed by \citet{Umeda03}. 
They assume that the matter in the mass coordinate between $M_{\rm cut}$ and $M_{\rm mix}$ are mixed at the explosion phase, and that certain fraction, $f_{\rm fb}$, of the mixed region is ejected whereas the rest of the mixed material falls back on to a black hole.  
They successfully reproduced the abundance pattern of an HMP star, HE0107-5240, using this model.  
The abundance patterns of other CEMP stars are also explained by employing other parameters for the mixing and fallback \citep{Umeda05, Iwamoto05, Tominaga14}.  

%(Binary scenario)
The second scenario is the binary scenario \citep{Suda04, Komiya07}, where binary mass transfer plays a role in the formation of CEMP-no stars as it does for \cemps\ stars.  
In the standard framework of the stellar evolution without extra mixing processes, EMP stars in the mass range of $\sim 0.8 \hyp 3.5 \msun$ undergo helium-flash driven deep-mixing (He-FDDM), which is triggered by the hydrogen engulfment by the helium flash convection, leading to the dredge up of carbon and \spr\ elements to their surface. 
For stars with $\gtrsim 3.5 \msun$, He-DDM does not take place.
The \spr\ nucleosynthesis in these stars depend on other sources of neutrons such as $^{22}$Ne burning.  
In any cases, \spr\ is not efficient and should be weakly enhanced even if carbon is dredged up by the third dredge-ups. 
The secondaries of these stars are expected to be CEMP-no stars through binary mass transfer events. 
\citet{Yamada19b} point out that secondary stars with lower mass ($\lesssim 3.5 \msun$) primaries with wider separations also have chance to be CEMP-no stars because barium abundance will not exceed $\abra{Ba}{Fe} = 0.5$ by a small amount of mass accretion. % when carbon abundance are $0.7 < \cfe < 1.1$. 
  
One of the criticisms for the binary scenario is on the fraction of stars with radial velocity variations among CEMP-no stars.
\citet{Starkenburg14} argues that the fraction of confirmed binarity is not significantly higher for CEMP-no stars than the fraction for stars without carbon enhancement. 
However, a small amount of wind mass transfer in a wide binary with period of $\sim 10000$ days or even larger is enough to enrich the secondary star to $\cfe > 0.7$, where the variations of radial velocities should be difficult to detect.   
\citet{Hansen16} reported binarity for three CEMP-no stars, all of which belong to the Group I.  
Their estimated binary periods fall in the same range as \cemps\ stars, which is consistent with the expectation that the Group II and Group III stars belong to long period binaries.
Indeed, one of the Group III stars, HE 0107-5240, was reported to have a companion star whose binary period is longer than 10000 days \citep{Arentsen19}.
We also note that some CEMP-no stars also show Eu/Ba ratio predicted by the \spr\ nucleosynthesis, and that the distribution of $\abra{Ba}{Fe}$ of CEMP-no stars is different from that of C-normal EMP stars \citep{Yamada19b}.  

%(Spinstar scenario)
The third scenario is the so called ``spinstar" scenario \citep{Meynet06, Maeder15}. 
In a fast-rotating metal-poor massive star, rotation-induced mixing takes place and enriches their surface with carbon and other light elements. 
The gas polluted by the carbon-enhanced wind from spinstars can be a progenitor of CEMP-no and CRUMP stars. 

%(Q)
All the three processes are asserted to produce carbon and other elements enhanced in CRUMP stars. 
However, the holistic distribution of the absolute abundances such as $\coh$ and $\feoh$ for CEMP stars is not well studied. 

%(analytic estimate)
In particular, there is a challenge to the faint SN scenario in terms of the formation of the subsequent star formation. 
The mass of carbon from a faint SN should be comparable to a normal SN ($\Delta M_{\rm C} \sim 0.2 \msun $), to obtain a large value of $\cfe$ from the small amount of ejected iron due to the large fraction of the fallback of the inner ejecta.  
If we assume that the carbon with mass $\Delta M _{\rm C}$ ejected by a first generation supernova is homogeneously mixed in the supernova remnant of mass $M_{\rm sw}$, the carbon abundance will be estimated as follows. 
\begin{equation}
\coh = -2.5 + \log \left[ \bfrac{\Delta M_{\rm C}}{0.1 \msun} \bfrac{M_{\rm sw}}{10^4 \msun}^{-1} \right] . 
\end{equation} 
This estimate indicates that the typical carbon abundance of second generation stars with a faint SN progenitor is lower than the carbon abundances of CRUMP stars with $ \coh \gtrsim -2 $. 
If we assume that the SN ejecta is mixed in a mini-halo with $\sim 10^{5}\msun$, the second generation stars of a normal SN progenitor will be born as EMP stars with $\feoh \simeq -4 \hyp -3$, and hence, with $\coh \simeq -4 \hyp -3$.  
For the case of faint SNe, they should produce the second generation stars in the similar range of $\coh \simeq -4 \hyp -3$ since $\Delta M_{\rm C}$ is comparable to that of normal SNe. 

%(Fe)
On the other hand, there is a challenge to the binary scenario and spinstar scenario on the origin of iron in CEMP-no stars. 
The binary mass transfer from AGB stars and wind from spinstars leave iron mostly untouched while carbon is enhanced.  
For CEMP-no stars with $\feoh \gtrsim -4$, the normal SNe can be the suppliers of iron. 
On the other hand, we need the explanation for the extremely low iron abundance in CRUMP stars. 
One possible scenario for a small amount of iron in CRUMP stars is the surface pollution after their birth \citep{Shigeyama03, Suda04, Komiya09L}. 
When low-mass Pop III stars are formed, their surfaces can be polluted up to $\feoh \sim -5$ by the accretion of interstellar medium (ISM) which is enriched with the metal ejecta of SNe, produced in the host mini-halos.
These stars are to be observed as HMP/UMP stars. 
Another possibility is the dilution of the ejecta from a normal SN in very large volume \citep{Karlsson06, Komiya15, Smith15}. 
A SN blast wave can blow-out the gas from their host mini-halo and enriches intergalactic medium (IGM) and other mini-halos. 
The gas enriched with the diluted SN ejecta may form HMP/UMP stars as second generation stars. 

%(this work)
In this paper, we examine the faint SN scenario in the framework of chemical evolution. 
We compute the chemical enrichment history of carbon and iron in the very early stages of the galaxy formation and compare with the observed EMP, UMP and HMP stars. 
The binary scenario is also examined by including the accretion of C-rich matter ejected from the envelope of AGB stars.
%(this work2)
%We also examine the binary + surface pollution scenario.  
We do not consider the spinstar scenario because there are subject to large uncertainties on the diffusion of stellar wind, the formation of next generation stars in the wind ejecta, and the final fate of spinstars after the wind mass loss.

%(previous chemical evolution)
The chemical evolution by the very early generations of stars, produced by faint SNe, has been investigated by some previous studies. 
\citet{Cooke14} modeled the enrichment of C and Fe by Pop III SNe in mini-halos to estimate the fraction of carbon enhanced stars using the SN model of \citet{Heger10}. 
They only considered the abundance distribution by a single star formation event at $z=20$ and ignored the influence of the formation history of galaxies.

\citet{Sarmento17} performed hydrodynamic simulations with a new subgrid model to predict the distribution of chemical composition of star particles. 
In their model, CRUMP stars with $\coh \simeq -2 \sim -1$ are formed. 
However, they assumed that all the Pop III stars have the yields taken from the best fit model for SMSS J031300.36-670839.3 by \citet{Heger16}.  
Also, their models are dependent on their sub-grid model of metal mixing, and are subject to uncertainties associated with the numerical scheme. 
\citet{Sharma16} discussed origin of CEMP stars using a cosmological hydrodynamical simulation. 
The drawback of their model is not to resolve the mass of of the mini-halos with $M_{\rm h} = 10^6 \msun$, and the estimate of the chemical composition of EMP stars is dependent on their sub-grid model. 
In all of the previous studies above, `star particles' do not mean individual stars but aggregations of stars. 
In this paper, we follow the formation and merging history of galaxies, together with the star formation history and the feedback effects in the host galaxies. 
We also take into account the pre-enrichment of IGM and the external-enrichment of proto-galaxies by the outflows driven by Pop III SNe. 
We treat the formation and explosion of individual metal poor stars in host galaxies, and compute the chemical composition of each low-mass star which is supposed to survive in the MW halo. 

%(organize)
This paper is organized as follows. 
In the next section, we describe our computational method, including new ingredients to investigate the contribution of faint SNe. 
In \S~\ref{S:result}, we present the results with the faint SN scenario, and their parameter dependence. 
In \S~\ref{S:binary}, we consider the contribution of binaries under the binary scenario. 
Conclusions follow in \S~\ref{S:conclusion}.

\section{Computational Method}\label{S:model}

%\subsection{StarTree code}
We have developed a chemical evolution model with merger trees in previous studies \citep{Komiya09L, Komiya10, Komiya14, Komiya16}, which we name the {\it StarTree} code. 
Here we only provide a brief summary of our code and describe what has been changed from our previous models and assumptions. 

%(inhomogeneous)
The main improvement of the {\it StarTree} code is the consideration of the chemical inhomogeneity of proto-galaxies in order to investigate the effect of faint SNe. 
We adopt a stochastic chemical evolution model by SNe, which is described in \S~\ref{S:stochastic}. 
The assumptions related to faint SNe are described in \S~\ref{S:faintSN}.  

%(tree, feedback)
We also updated the method of building merger trees (\S~\ref{S:tree}) and the schemes for the radiation feedback on Pop III star formation (\S~\ref{S:feedback}). 

%(no pollution)
As mentioned above, the surface pollution of stars by ISM accretion can be important for HMP/UMP stars but we do not consider the surface pollution in discussing the faint-SN scenario. 
This is because the abundance patterns of the elements up to the iron group elements are explained by faint SNe without surface pollution for CEMP-no or CRUMP stars.  
Although our code can compute the surface pollution of Pop III or EMP stars, we switched off the pollution subroutine in the following computations for faint SNe. 

%(binary)
We also consider binary mass transfers as an alternative scenario for CEMP-no stars (\S~\ref{S:binary}). 
The assumptions about the binary mass transfer events are described in \S~\ref{S:binary}.

\subsection{The StarTree code}\label{S:StarTree}

We build merger trees of galaxies based on the extended Press-Schechter method, and follow the chemical enrichment along the trees. 
We refer the baryonic components in the mini-halos as proto-galaxies. 

%(SFR)
We assume that the star formation rate, $\dot{M_*}$, in a proto-galaxy is proportional to gas mass $M_{\rm gas}$,
\begin{equation}
\dot{M_*} = \epsilon_{\star} M_{\rm gas}. 
\label{eq:sfr}
\end{equation}
where $\epsilon_{\star}$ is the star formation efficiency, depending on the mass of the host halo, and is defined by
\begin{equation}
\epsilon_{\star} = 1.2 \times 10^{-14} (M_{\rm h}/\msun)^{0.3} \hbox{ yr}^{-1}.
\end{equation}
The fiducial value is chosen to reproduce the observed mass-metallicity relation for dwarf galaxies, and to be consistent with the abundance distribution of {\it r}-process elements in EMP stars under the assumption that their sources are the ejecta of coalescing neutron star binaries \citep{Komiya16b}.

%(individual star)
A central feature of the {\it StarTree} code is that all the individual Pop III and EMP stars are registered. 
We consider each massive star and explore the properties of low-mass survivors.  
We randomly set the mass of each star according to the adopted IMF. 
We use the lognormal form of the IMF with the modification by the power-law tail at higher mass following \citet{Chabrier03}. 
The binary fraction is set at $f_b = 50$ \%, where the mass of the primary stars are subject to the same IMF as single stars, while the mass of the secondary stars are determined by the mass ratio function.
We employed the flat mass-ratio distribution as in the previous studies \citep{Komiya07,Komiya09a}.
The peak mass, $\mmd$, of the IMF for EMP and Pop III stars is set at $3$ and $10\msun$, respectively, to be consistent with the fraction of known \cemps\ stars as discussed in our previous studies \citep{Komiya07, Suda13, Komiya16b}. 

%(outflow, infall)
We also considered the gas infall onto the mini-halos, the gas outflow by SN explosion, and the pre-enrichment of IGM by the gas outflow. 
The gas infall rate is assumed to be proportional to the dark-matter infall rate given by the Press-Schechter merger tree. 
We computed the mass and metal outflow rates by individual SNe as a function of the SN explosion energy and the binding energy of proto-galaxy (see also \S~\ref{S:infall}). 
We followed the evolution of the winds from proto-galaxies by assuming the momentum-conserved snowplow model, and considered the inhomogeneous metal enrichment in the IGM. 
We evaluated the distance between two proto-galaxies in the merger trees,
which were used to estimate the amount of pre-enrichment in newly formed proto-galaxies and of external enrichment from other existent proto-galaxies. 

%()
We used the same assumptions and model parameters as in \citet{Komiya16b} otherwise described here and in the following.

\subsubsection{Merger Tree}\label{S:tree}

We have updated the realizations of merger trees by replacing the method of \citet{SK99} with that of \citet{Parkinson08}. 

It is pointed out that the Press-Schechter mass function based on the spherical collapse model overestimates the number of small halos \citep{Zhang08} compared with the results of N-body simulations and the results with the elliptical collapse model \citep{Sheth01, Sheth02}. 
\citet{Parkinson08} modified the mass distribution function given by the Press-Schechter theory and presented a method to build merger trees by which the results of the N-body simulations are well reproduced. 

%(new tree)
The modification of the {\it StarTree} code resulted in the reduction in the number of the branches of merger trees.  
The number of the branches of the trees to produce MW mass with the mass of $M = 10^{12} \msun$ has decreased to $\sim$ 60,000 in the new model, from 200,000 $\hyp$ 300,000 in the previous models.  
The smaller the number of branches the smaller the number of Pop III and EMP stars, while the predicted abundance distributions in the following sections are not significantly affected, in particular after the typical mass of mini-halos have $M_{\rm h} \gtrsim 10^7 \msun$.  %%% ???(MYF)

%(cosmological parameters)
Cosmological parameters were updated to those provided by \citet{Planck16}.

\subsubsection{Suppression of Pop III Star Formation by UV Radiation}\label{S:feedback}
%(previous)

Pop III star formation in mini-halos is thought to be suppressed by the irradiation of the Lyman-Werner (LW) photons. 
To account for this effect, we assumed that Pop III proto-galaxies with the virial temperature lower than $10^4$ K do not form stars below the critical redshift $z_{\rm LW} = 20$ \citep{Komiya16b}. 
We have improved the prescription for the effect of the LW photos by computing the flux of the LW radiation and the suppression of star formation in a self-consistent way in each time step.
This new treatment comes from our consideration that the LW radiation depends on a star formation rate and the IMF of Pop III stars.

%(new method)
\citet{Machacek01} provided the critical mass, $M_{\rm crit}(F_{\rm LW})$, below which the Pop III mini-halos do not form stars, as a function of the LW luminosity, $F_{\rm LW}$. 
\citet{Oshea08} modified the critical mass by a factor of four based on their cosmological simulations. 
We adopt their formula, 
\begin{eqnarray}
& & M_{\rm crit}(F_{\rm LW}) = 4 \times \Bigl [1.25 \times 10^5 \msun  \Bigr. \nonumber \\
& & \qquad + 8.7 \times 10^5 \msun \bfrac{F_{\rm LW}}{10^{-21} {\rm erg^{-1} cm^{-2} Hz^{-1}}}^{0.47} \Bigl. \Bigr] . 
\end{eqnarray} 
Using the emission rate of the LW photons from Pop III stars by \citet{Schaerer02}, we computed $F_{\rm LW}$ in each mini-halo by summing up the radiations from other mini-halos. 

%()
The suppression by the LW radiation works only for Pop III proto-galaxies and is not applied to pre-enriched proto-galaxies with mass below $M_{\rm crit}$ due to the cooling by metallic lines.

\subsection{Stochastic Chemical Enrichment in a Proto-Galaxy}\label{S:stochastic}

\begin{figure}
\includegraphics[width=\columnwidth,pagebox=cropbox]{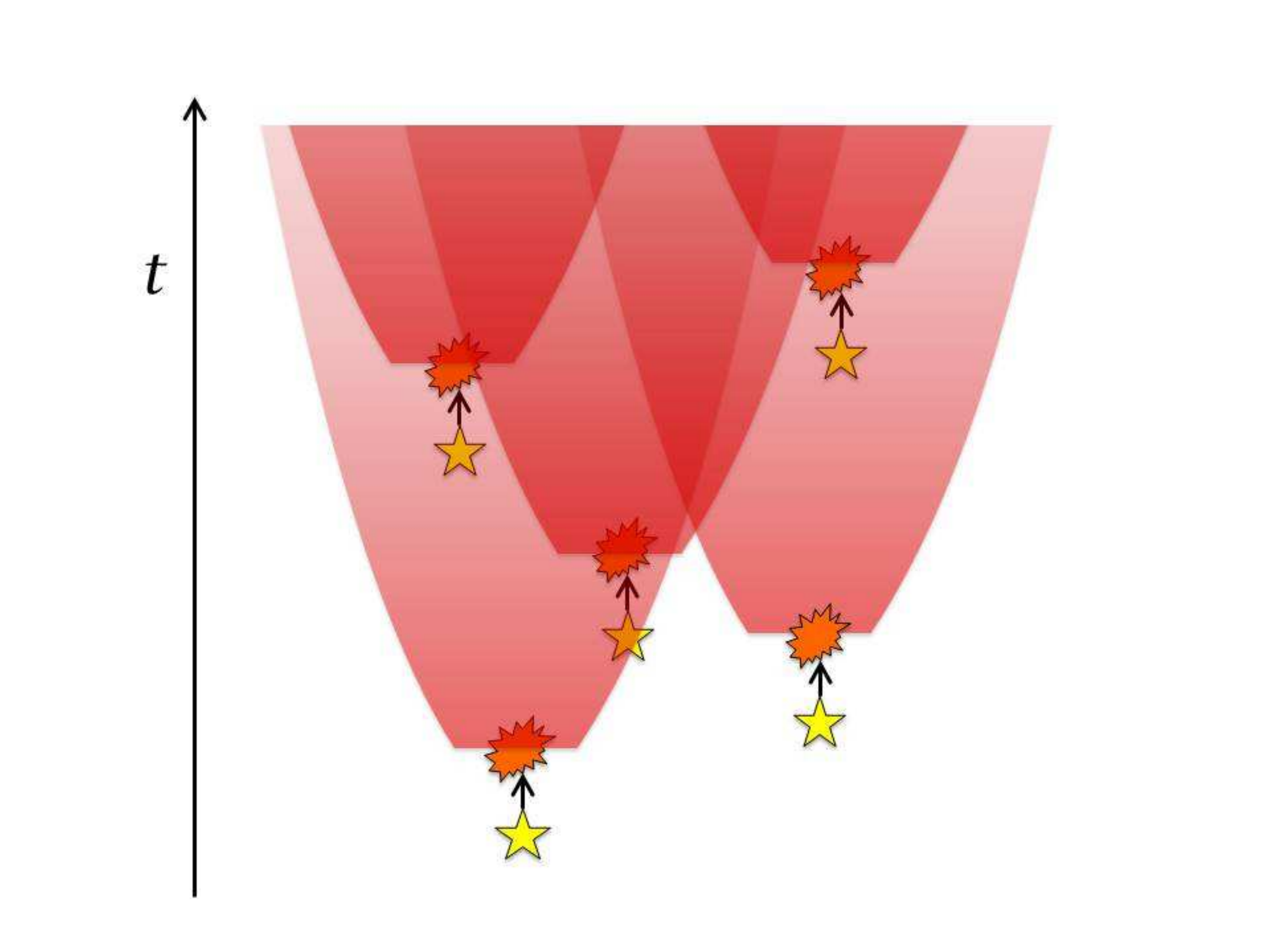}
\caption{
A schematic illustration of our stochastic chemical evolution model. 
The star symbols in yellow and the star polygons in red represent massive stars and the end of their lives as supernovae, respectively.
The red shaded areas are the ISM polluted by the yields of SNe, which expands in time by the diffusion process.
The chemical enrichment of the region is affected by the sum of the pollution from SNe. 
}\label{cartoon}
\end{figure}

We modeled the inhomogeneous chemical evolution in a proto-galaxy as a stochastic process of metal enrichment by a number of SNe. 
Figures~\ref{cartoon} and \ref{cartoon2} give schematic views on the stochastic chemical enrichment model of the mini-halos. 
The assumptions and numerical parameters are described in the following subsections.

As shown in Fig.~{\ref{cartoon}, a SN explosion produces a metal-polluted region. 
The polluted region evolves with time by the diffusion process if the ejected matter remains in the host halo. 
The accumulation of SN yields is built up through the overwrapping of polluted regions. 

Some of the outflow gas and metals, which are triggered by SNe, fall back again onto their host halos or other nearby halos, along with the infall of IGM gas. 
The metal abundances of the infall-gas are recorded and are traced in each tree (see \S~\ref{S:infall}). 
   The gas expelled as a galactic wind is assumed to form a spherical shell around the mini-halos, i.e., the accreted gas does not mix with the host halo.
The ejected gas is the source of the next generation stars.
The star formation rate is in proportion to the mass of the gas in ISM, regardless of the metallicity or the density of ISM. 
We do not consider the star formation triggered by the SN.

\begin{figure}
\includegraphics[width=\columnwidth,pagebox=cropbox]{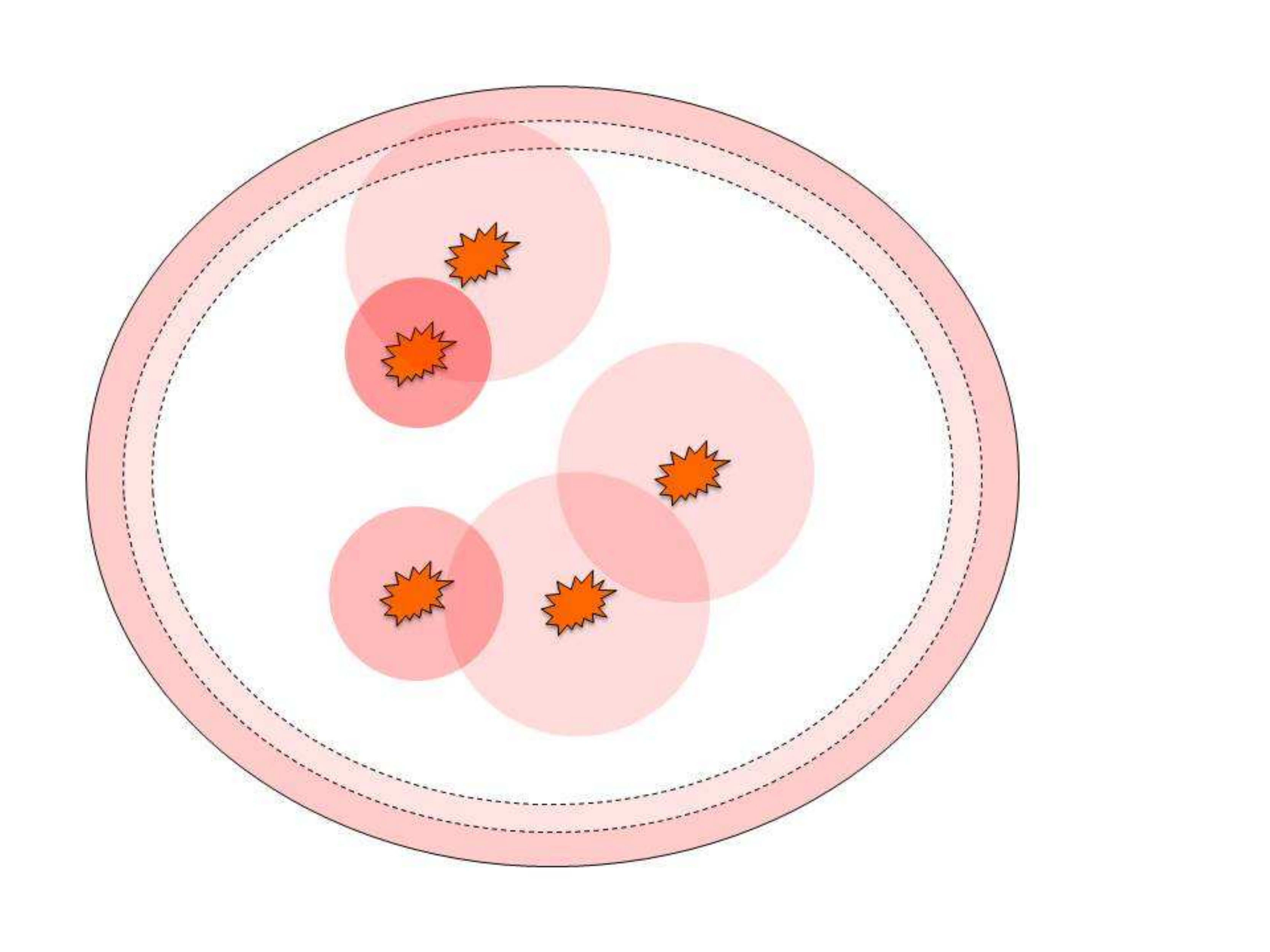}
\caption{
A schematic illustration of the metal enrichment in a proto-galaxy.  
The star polygons in red represent the SNe in the host halo whose boundary is shown by the ellipse in the solid line.
The red shaded areas are the ISM influenced by the SN yields.
The inner ring in thin red stands for the re-accretion of the outflow gas by SNe, where we assumed that the part of the ejected gas from the host halo accretes on itself again.
The outer ring in dense red corresponds to infalling gas from other mini halos, which works as the pre-enrichment of the halo with metals.
These rings show the initial pollution process in a simplified way.
In our simulations, the accreted gas retains in the outer part of the host halo, and contributes to the abundances of the subsequent generations of stars.
See text for more details on the accretion and mixing of the SN yields.
}\label{cartoon2}
\end{figure}

\subsubsection{Metal Pollution by each Supernova}

%(M_p initial)
The initial mass of the polluted region, $M_{{\rm p},i}$, occupied by the $i$-th SN is given by
\begin{equation}\label{eq:mpidef}
M_{{\rm p},i} =  M_{{\rm sw}, i} - M_{{\rm w}, i}  
\end{equation}
where $M_{{\rm sw}, i}$ and $M_{{\rm w}, i}$ denotes the swept-up mass and the mass of the galactic wind (or outflow) ejected from the proto-galaxy by the $i$-th SN, respectively.  
The swept-up mass is prescribed as follows according to \citet{Shigeyama98}; 
\begin{eqnarray}
M_{\rm sw} = 5.1 & \times & 10^4 \msun \bfrac{E_{\rm SN}}{10^{51} {\rm erg}}^{0.97}  \nonumber \\
& \times & \bfrac{n_1}{1 {\rm cm}^{-3}}^{-0.062} \bfrac{c_s}{10 \, {\rm km \, s}^{-1}}^{-9/7}, 
\label{eq:M_SNR}
\end{eqnarray}
where $E_{\rm SN}$ is the explosion energy of the SN, $n_1$ the number density of interstellar gas, and $c_s$ the sound speed.    
The dependences on density and the sound speed are dropped for simplicity in our models.
Therefore, the above equation reduces to
\begin{equation}\label{eq:M_sw51}
M_{\rm sw} = M_{\rm sw51} \bfrac{E_{\rm SN}}{10^{51} {\rm erg}}^{0.97} 
\end{equation}  
where $M_{\rm sw51} = 5.1 \times 10^4 \msun $ is our fiducial value. 
We also tested a smaller value by a factor of ten, i.e., $ M_{\rm sw51} = 5.1 \times 10^3 \msun $.
This is intended to form second generation stars with larger value of $\coh$ from smaller swept-up mass by the first generation SN.
However, we note that the assumption of a small swept-up mass is likely to be implausible for the formation of second generation stars. 
First, as seen from eq.~(\ref{eq:M_SNR}), the weak density dependence of swept-up mass does not allow us to assume high density in the ISM. 
Second, it is also difficult to assume large sound speed in the host halo because the fiducial value of $c_s = 10 {\rm km/s}$ corresponds to $T = 10^4 {\rm K}$ and is comparable to the escape velocity of mini-halos hosting Pop III stars. The assumption of a larger value for $c_s$ is unrealistic in terms of star formation.

The ejected mass by the galactic wind, $M_{\rm w}$, is taken from the prescription in \citet{Komiya14} as a function of $E_{\rm SN}$ and the binding energy, $E_{\rm bin}$, of the proto-galaxy.   
The energy and the mass load of the galactic wind are given by the following interpolation formulae;
\begin{eqnarray}
E_{\rm w} & = & \eta E_{\rm SN} \bfrac{\epsilon + \eta E_{\rm SN} /E_{\rm bin}}{1 + \eta E_{\rm SN} /E_{\rm bin}} 
\label{eq:windenegy} \\
M_{\rm w} & = &M_{\rm gas} \bfrac{E_{\rm w}}{E_{\rm bin}}, 
\label{eq:windmass}
\end{eqnarray}  
   where $\eta$ and $\epsilon$ is the fraction of SN explosion energy converted into the kinetic energy of the gas shells, and the fraction of the kinetic energy converted into the wind energy, respectively.       
The definition of $M_{\rm gas}$ is the same as in eq. (\ref{eq:sfr}).
   The ejected mas approaches $M_{\rm w} = M_{\rm gas}$ if $\eta E_{\rm SN} \gg E_{\rm bin}$ and $M_{\rm w} = \eta \epsilon M_{\rm gas} ({E_{\rm SN}}/{E_{\rm bin}})$ if $\eta E_{\rm SN} \ll E_{\rm bin}$. 
   We adopt fixed values of $\eta = 0.1$ and $\epsilon = 0.1$ \citep[see][]{Komiya14}.
   If $M_{\rm w}$ exceeds $M_{\rm sw}$, we presume that all the swept-up matter is blown away, i.e., $M_{\rm w} = M_{\rm sw}$ and $M_{\rm p} = 0$. 

%(dM_p/dt)
The evolution of the mass of the metal-polluted region by the $i$-th SN is described as follows, 
\begin{eqnarray}\label{eq:dmpidt}
\frac{ d M_{{\rm p}, i} }{dt} = & C_{\rm diff} (t-t_i)^{1/2} - \sum_{j > i} \delta({t-t_j}) M_{{\rm w},j} \bfrac{M_{{\rm p}, i}}{M_{\rm gas}} & \nonumber \\
& - \sum_{\{k|i \in S_k\}} \delta({t-t_k}) m_{*,k} .  &
\end{eqnarray}
Here $t_i$ is the time of the $i$-th SN explosion and $m_{*,k}$ is the mass of the $k$-th star in the proto-galaxy:  
The first term in the right hand side describes the diffusion of the polluted gas.   
The diffusion equation in a constant density gives the time dependence of $ M \propto t^{3/2} $, and hence, $dM/dt \propto t^{1/2}$:  
The second and third term corresponds to the reduction of the polluted mass by the galactic wind driven by later SNe and by the formation of the next generation stars, respectively: 
$S_k$ is a set of the polluted regions that contribute to the formation of the $k$-th stars (see \S~\ref{S:starform}). 
If the mass of the polluted region increases to the gas mass of the proto-galaxy, it stops growing and keeps the value of $M_{{\rm p}, i} = M_{\rm gas}$ accordingly. 

The diffusion coefficient, $C_{\rm diff}$, which is a free parameter, is set at $10^{-6} \msun \, {\rm yr}^{-3/2}$ as the fiducial value. 
This corresponds to a reasonable environment for the formation of first stars in proto-galaxies, where the gas of $\sim 2\times10^5 \msun$ is mixed in a dynamical timescale ($\sim 3 \times 10^7$ yrs) for the virialized halo at $z=10$. 
Considering the turbulent mixing with the velocity $v_{\rm turb}$ and the scale-length $l_{\rm turb}$, we obtain
\begin{eqnarray}
C_{\rm diff} &=& \frac{4\pi}{3} \bfrac{v_{\rm turb} l_{\rm turb}}{3}^{3/2} \mu n m_{\rm p}  \\
&=& 6 \times 10^{-7} \msun \, {\rm yr}^{-3/2} \bfrac{v_{\rm turb}}{10 \hbox{ km s}}^{3/2} \bfrac{l_{\rm turb}}{100 \rm pc}^{3/2} n_1 \mu, \nonumber
\end{eqnarray}
where $\mu$ is the mean molecular weight and $m_{\rm p}$ the proton mass. 

%(A_Z initial)
The mass of element species Z in the $i$-th polluted region should have an initial value of 
\begin{equation}
A_{{\rm Z}, i} = (1-f_{{\rm w},i}) Y_{{\rm Z},i}
\end{equation}
where $Y_{{\rm Z}, i}$ is the SN yield of element Z in mass, and $f_{{\rm w}}$ is the metal loading factor of the SN-driven galactic wind.  
The value of $f_{{\rm w}}$ is taken from the formula in \citet{Komiya14};  
\begin{equation} 
f_{\rm w} = \min \left(1, \frac{M_{\rm w}/M_{\rm sw} + \eta E_{\rm SN} / E_{\rm bin}}{1 + \eta E_{\rm SN} / E_{\rm bin}}\right).
\label{eq:fracwind}
\end{equation}   
It approaches $M_{\rm w}/M_{\rm sw}$ if $\eta E_{\rm SN} \ll E_{\rm bin}$, and $f_{\rm w} = 1$ at the opposite limit. 

%(dA/dt)
The formulation of the time evolution of the mass of metals, $A_{{\rm Z}, i}$, can be written in analogy to eq.~(\ref{eq:dmpidt}).
The mass of the metals in the $i$-th polluted region, $A_{{\rm Z}, i}$, is conserved through the diffusion. 
The outflow and star formation do not change the chemical composition ${A_{{\rm Z}, i}} / {M_{{\rm p}, i}}$.  
Therefore, the change of the mass of the metals with respect to time is given as follows.
\begin{eqnarray}
\frac{ d A_{{\rm Z}, i} }{dt} & = & - \sum_j \delta({t-t_j}) A_{{\rm Z}, i} \frac{ M_{{\rm w}, j}}{M_{\rm gas}} \nonumber \\
& & - \sum_{\{k|i \in S_k\}} \delta({t-t_k}) A_{{\rm Z}, i} \frac{ m_{*,k}}{M_{\rm p, i}} 
\end{eqnarray}

%(mix)
To save a computational cost in computing inhomogeneous chemical evolution for EMP stars, we introduce a cutoff parameter for chemical enrichment.
The inhomogeneous mixing by SN ejecta is switched off if the number of polluter SNe in a proto-galaxy exceeds the critical value.
We treat the mixing of metals in a proto-galaxy as the averaged pollution if the number of SNe is more than 30, i.e.,
\begin{equation}
\sum_i M_{{\rm p}, i} = 30 M_{\rm gas}, 
\label{eq:30mix}
\end{equation}
or if the average metallicity of the proto-galaxy exceeds $\feoh=-1$.  
The fraction of gas that remains unpolluted by SNe is evaluated by $\prod_i ( 1-{M_{{\rm p},i}}/{M_{\rm gas}} )$. 
If all the polluted regions have the same mass of $M_{\rm p} (< M_{\rm gas})$, this value is reduced to $( 1 - {M_{\rm p}}/{M_{\rm gas}} ) ^ {30 M_{\rm gas}/ M_{\rm p}} $, which is below $10^{-13}$ for any ratios of ${M_{\rm p}}/{M_{\rm gas}}$. 
For any choice of $M_{\rm p}$, the fraction is always small. 
The application of this mixing typically occurs at $\feoh \sim -3$ in the fiducial model.
We checked the validity of this model by employing a larger value for the criterion of $\sum_i M_{{\rm p}, i} = 60 M_{\rm gas}$, which gives an almost identical result.

\subsubsection{Infall of Metal-Enriched Gas}\label{S:infall}

We consider the gas infall onto proto-galaxies in addition to the metal-enrichment of IGM by the galactic wind. 

As mentioned above, the first SN in a proto-galaxy triggers the galactic wind into the IGM. 
As the proto-galaxy evolves, metals and gas blown-out by subsequent SNe are added to the wind. 
We follow the spread of the wind in the IGM assuming momentum conservation (see \citet{Komiya14} for details). 

The mini-halos accrete IGM in accordance with the growth of the mini-halo. 
Though the IGM has the pristine abundance at very early universe, the IGM around the mini-halo is enriched by the wind from nearby proto-galaxies and/or the wind from the mini-halo itself. 
Therefore, the chemical abundance, $ X_{{\rm inf}, Z} $, of the infalling gas depends on time. 
The chemical composition of a proto-galaxy can be inhomogeneous even before the first SN in the proto-galaxy. 

Following our previous studies, we assume that the gas infall rate is proportional to the mass evolution of the dark halo given by a merger tree.  
%(inhomogeneous)
We store the data of the metallicity, $ X_{{\rm inf}, {\rm Z}}(z) $, and the mass, $\Delta M_{\rm gas}(z)$, of the infall-gas budget for each timestep, $\Delta z$, and each branch of the merger trees. 
The timestep is set at $\Delta z = 0.01 (1+z)$. 
The diffusion of metals in the ejected and infalling gas is not taken into consideration.

\subsubsection{Stellar Abundances}\label{S:starform}

The chemical abundance of stars is given by the sum of the elements for metals, Z, carried by the infalling gas, and taken from the metal-polluted regions by SNe. 
When a $k$-th star is formed, we randomly select an infall-gas budget with the probability proportional to the mass, $\Delta M_{\rm gas}(z)$, of each gas budget.  
Then, we randomly determine whether the star is in the $i$-th polluted region or not (i.e., $i \in S_k$ or not) by the probability of $M_{{\rm p}, i} / M_{\rm gas}$, for each of all the polluted regions by SNe.  
We compute the sum of the chemical abundance of the selected infall-gas budget and the polluted region, in which the star is formed, 
\begin{equation}
X_{k, {\rm Z}} = X_{{\rm inf},{\rm Z}} + \sum_{\{k|i \in S_k\}} \frac{A_{{\rm Z}, i}}{M_{{\rm p}, i}}, 
\end{equation}
and record this value as the abundance of $k$-th star.

\subsection{Faint Supernovae}\label{S:faintSN}

The yields of faint SN are determined by a free parameter to produce a wide range of the observed values of $\cfe$ in CEMP-no stars.
We introduced a fallback parameter $f_{\rm fb}$ to control the value of \cfe\ by decreasing the iron yield by a factor of $f_{\rm fb}$, keeping the carbon yield unchanged.
We adopted chemical yields by \citet{Kobayashi06} where the systematic mass dependence of the yields from normal SNe is available.
The choice of the parameter from $0.1$ to $10^{-5}$ produces the observed range of $1 \lesssim \cfe \lesssim 5$ for CEMP-no stars.
The values of $f_{\rm fb}$ are subject to the log-flat distribution in the range of $0.1 \hyp 10^{-5}$.
Comparisons are made only for carbon and iron abundances in this study.

%(Z_cr)
The occurrence of faint SNe are also determined by a free parameter due to our poor knowledge on the formation of SNe, i.e., we do not know the ratio of faint SNe to normal SNe, and how it depends on mass and metallicity.
We tested the case with the most efficient contribution of faint SNe by introducing the critical metallicity, $Z_{\rm cr}$ below which all the massive stars end their lives as faint SNe.
The fiducial value of $Z_{\rm cr} = 10^{-5} \Zsun$ is adopted.
The results with the other choice of this parameter are discussed in Section~\ref{S:Zcr}

The key parameters in the {\it StarTree} code are provided in Table~\ref{params}.
The second, third, and fourth column represents the meaning of the parameter, the fiducial value, and other values to examine parameter dependence, respectively.

\begin{table*}
\begin{center}
\caption{Key parameters and values in the {\it StarTree} code}
\label{params}
\begin{tabular}{llcc}
\hline
parameter & description & fiducial value & other tested values \\
\hline
$\epsilon_{\star}$     &  star formation efficiency        &  $1.2 \times 10^{-14} {\rm yr}^{-1}$  &  $5 \times 10^{-11} {\rm yr}^{-1}$   \\
$\dot{M_*}$            &  star formation rate              &  $\epsilon_{\star} M_{\rm gas}$       &  n/a  \\
$f_{b}$                &  binary fraction                  &  50 \%                                &  n/a  \\
$M_{\rm md}$           &  Peak mass of the IMF             & $10\msun$ for Pop.~III and $3 \msun$ for EMP stars  & $100 \msun$ for Pop.~III  \\
$M_{{\rm sw},i}$       &  swept-up mass by the $i$-th SN  &  $5.1 \times 10^{4} \msun$  &  $5.1 \times 10^{-3} \msun$ \\
$M_{{\rm w},i}$        &  mass of the galactic wind triggered by the $i$-th SN      &  n/a$^{a}$                  &  n/a  \\
$M_{{\rm p},i}$        &  mass of the polluted region occupied by the $i$-th SN  &  eq.~\ref{eq:mpidef} and eq.~\ref{eq:dmpidt}  &  n/a  \\
$E_{\rm SN}$           &  explosion energy of the SN       &  $10^{51}$ erg             &  n/a  \\
$E_{\rm bin}$          &  binding energy of the proto-galaxy  &  n/a$^{a}$              &  n/a  \\
$C_{\rm diff}$         &  diffusion coefficient for ISM mixing by SN ejecta  &  $10^{-6} M_{\odot} {\rm yr}^{-3/2}$              &  $10^{-5} M_{\odot} {\rm yr}^{-3/2}$, $10^{-7} M_{\odot} {\rm yr}^{-3/2}$  \\
$f_{\rm w}$            &  metal loading factor of the SN-driven galactic wind  &  eq.~\ref{eq:fracwind}              &  n/a  \\
$f_{\rm fb}$           &  fallback parameter of faint SNe  &  log-flat distribution in $0.1 \hyp 10^{-5}$              &  n/a  \\
$Z_{\rm cr}$           &  critical metallicity for faint SNe  &  $10^{-5} Z_{\odot}$    &  $10^{-3} Z_{\odot}$  \\
\hline
\end{tabular}
\tablenotemark{a}{The values are computed internally. See text for details.}
\end{center}
\end{table*}

\section{Results and Discussion}\label{S:result}

\subsection{Fiducial Model}\label{S:faintResult}

\begin{figure}
\includegraphics[width=\columnwidth,pagebox=cropbox]{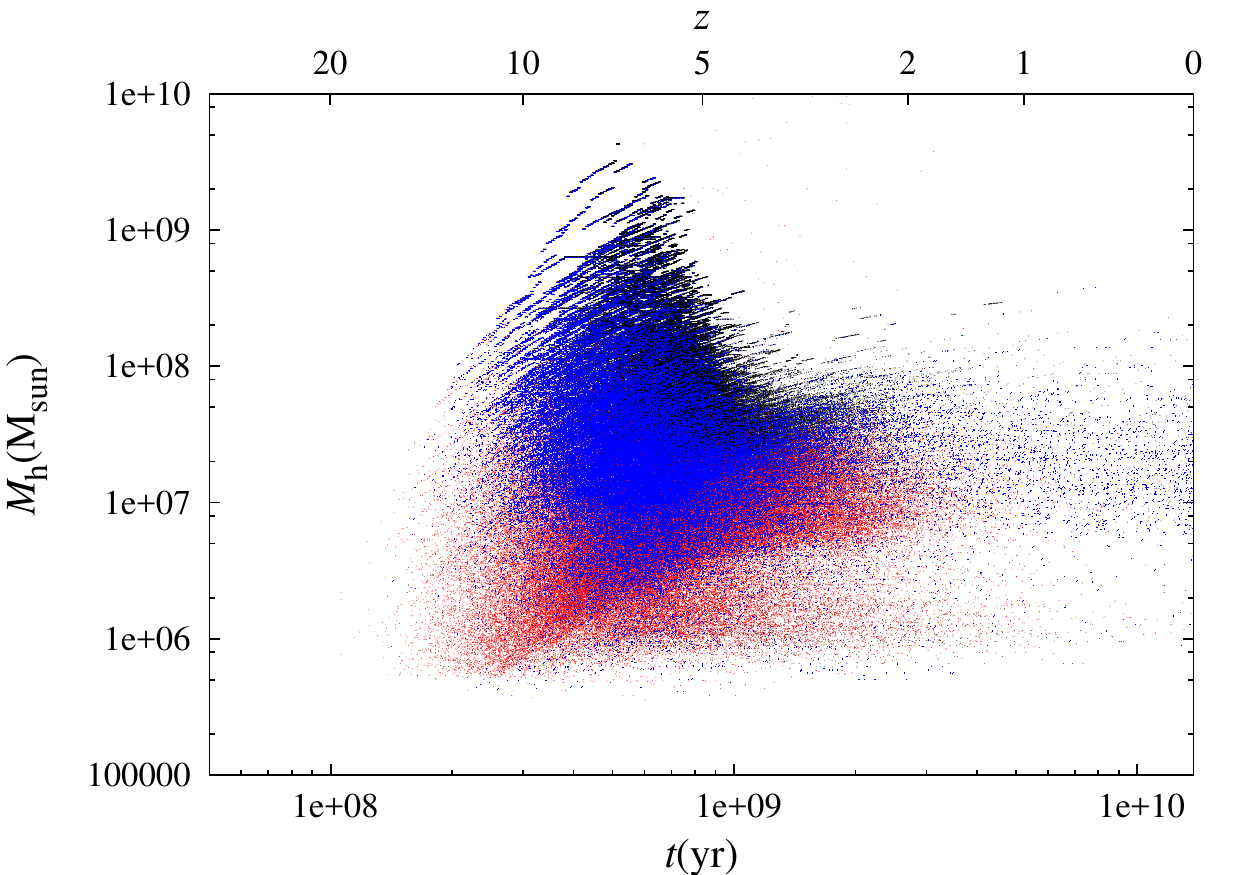}
\caption{
Time from the Big-Bang (abscissa) and the mass of the host mini-halos (ordinate) of faint SNe (red).  
Also plotted are the formation time (or redshift on the top) and the host halo mass of CEMP stars (blue) and EMP stars (gray) for $\feoh < -3$, which mostly overlap with the data points of faint SNe except for the later stages of evolution with larger mass of host halos. 
}\label{z-m}
\end{figure}

%(formation history)
Our fiducial model produced $\sim$160,000 faint SNe, where we found $2 \hyp 3$ faint SNe per branch of the merger tree.  
Most of their progenitors are Pop III stars, with the minor contribution of $\sim$ 15,000 second-generation stars having $Z \leq Z_{\rm cr}$. 

Figure~\ref{z-m} shows the explosion epoch and the mass of their host mini-halos for faint SNe (red), low-mass CEMP stars (blue), and EMP stars (gray). 
In a mini-halo of mass with $M_{\rm h} \lesssim 10^6 \msun$, faint or normal SNe blow off most of the gas from the host-halo. 
Even in this case, there is still a channel of CEMP star formation if the mini-halo attains the blown-off gas by fallback.
A majority of faint SNe takes place in mini-halos with the mass $M_{\rm h} < 10^7 \msun$, while only $15 \%$ of CEMP stars are formed in mini-halos in this mass range. 
Approximately $\sim 70 \%$ of CEMP stars are formed in halos with the mass of $10^7$ to $10^8 \msun$.

\begin{figure}
\includegraphics[width=\columnwidth,pagebox=cropbox]{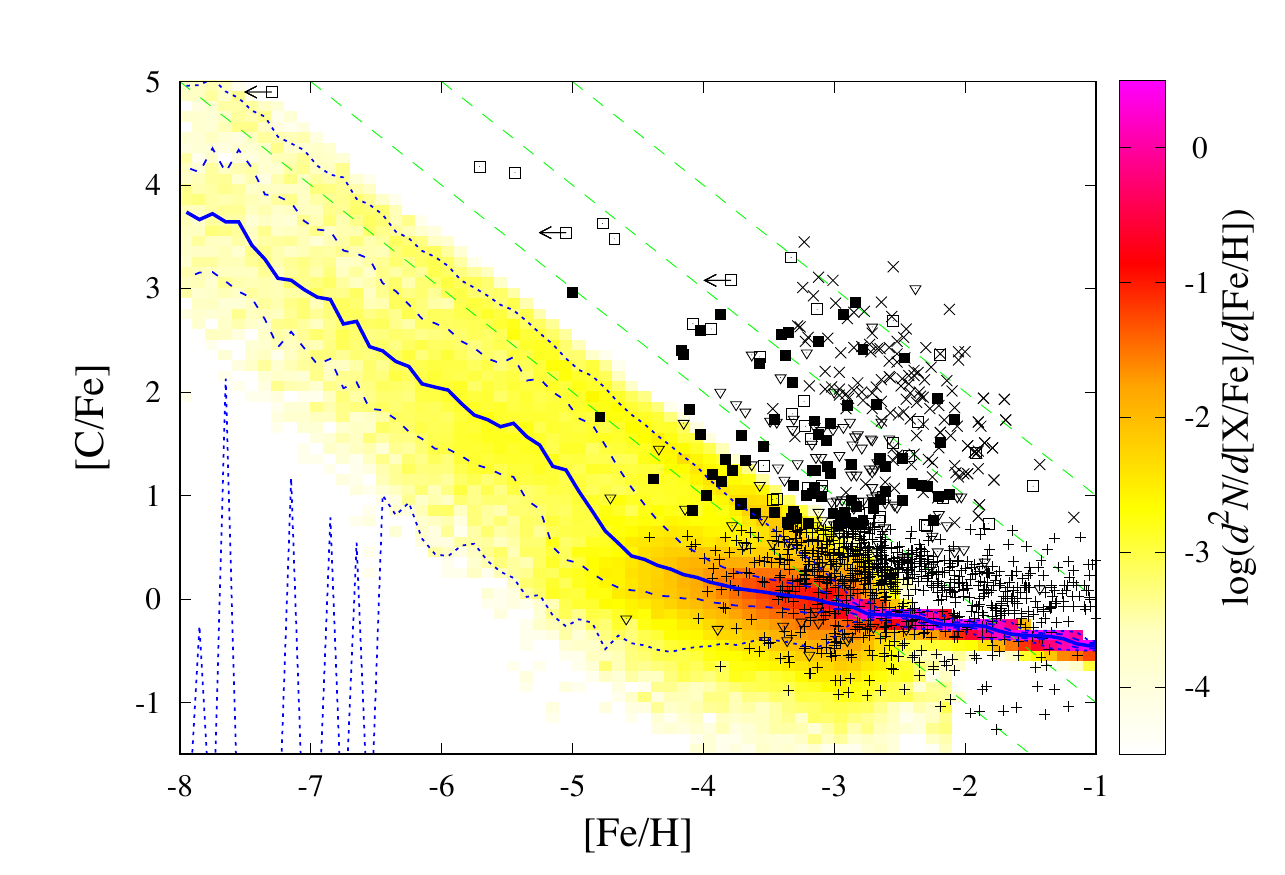}
\caption{
Distribution of the metal-poor stars on the metallicity ($\feoh$) and the carbon enhancement ($\cfe$) diagram. 
The predicted distribution is color coded as illustrated in the right margin. 
Blue lines show the percentile curves of $5, 25, 50, 75$, and $95 \%$ in each metallicity bin for the predicted distribution. 
Black symbols denote the observed data of stars, taken from the SAGA database, which have the same meanings as in Fig.~\ref{obs}. 
Diagonal dashed (green) lines indicate the loci of constant carbon abundances of $\coh = 0, -1, -2, -3$ from top to bottom, respectively. 
}\label{fiducial}
\end{figure}

%(CEMP)
Figures~\ref{fiducial} and \ref{MDFfid} show the distribution of the iron and carbon abundances and the metallicity distribution function (MDF) for the low-mass stars that survive to date, respectively. 
In Fig.~ \ref{fiducial}, the HMP and lower metallicity stars ($\feoh \le -5$) in the model show carbon enhancement ($\cfe \ge 0.7$), and yet, the carbon abundance is confined below $\coh \lesssim -2.5$, significantly lower than observed from CRUMP stars.  
For UMP stars of $-5 < \feoh \le -4$, a majority of stars are predicted below $\cfe = 0.7$ and scarcely show carbon enhancement.  
For higher metallicity, only a small fraction of stars show carbon enhancement while their carbon abundances increase with iron abundances.  
The majority of stars with the largest carbon abundances are originated from the fallback gas, i.e., ejected from host halo by the SN-driven galactic wind, and then, re-accreted onto themselves.  

%(C/H upper limit)
There are no stars distributed in the large carbon abundance above $\coh > -2.5$ in this model for $\feoh \lesssim -3$. 
This is because the carbon abundance of a polluted region just after a faint-SN explosion is determined by 
\begin{equation}\label{eq:CoHup}
\coh \simeq \log \bfrac{\Delta M_{\rm C} / X_{\rm C \odot}}{M_{\rm sw} X_{\rm H, gas}/ X_{\rm H \odot}} \sim -2.5, 
\end{equation}
from the carbon budget ($\Delta M_{\rm C} = 0.2 \msun$) and the swept-up mass ($ M_{\rm sw} = M_{\rm ws51}$) of faint SNe.  
This sets an upper limit on $\coh$ of the second generation stars.  
The location of $\feoh$ of the second generation stars on the line of constant \coh\ depends on the fallback parameter $f_{\rm fb}$ in faint SNe, which controls the carbon enhancement, $\cfe$. 

%(diffusion)
In the polluted region by the accretion of IGM gas, the carbon and iron abundances are further reduced by mixing with the ISM through the diffusion, where stars with lower $\coh$ are formed. 
Since both iron and carbon are diluted in the same way, the abundance ratios of carbon to iron will not change by the diffusion.
In such a case, the stars move horizontally to the left on the $\feoh$ and $\cfe$ plane. 
CEMP stars are formed in the proto-galaxies with typical gas mass of $M_{\rm gas} \simeq 10^6 \, \hyp 10^7 \msun$, and are distributed around $\coh \simeq -4$. 

%(2nd generation faint SN)
A minor population of second-generation stars having $Z < Z_{\rm cr}$ are formed out of the first SN ejecta mixed with the ISM of mass larger than $\sim 10^7 \msun$.
They comprise only $\sim 2 \%$ of the total number of faint SNe in the fiducial model. 
%(C normal)
On the other hand, a majority of stars with $\feoh > -5$ are formed around $\coh \simeq 0$.  
These carbon-normal stars are the third or later generations of stars and are formed of gas polluted by a single or multiple normal SNe. 

%(discussion)
The fiducial model fails to explain the stars with a large value of $\coh$ among observed CRUMP stars, even if we consider the uncertainties of our models associated with the assumptions. 
For instance, it is possible to reproduce large carbon abundances by considering chemical inohomoeneity in mini-halos, caused by the pollution of a single faint SN. 
However, the swept-up mass and the diffusion coefficient of SN yields have to be smaller by more than an order of magnitude than our fiducial value because typical carbon abundances of the second generation stars are significantly lower than the observed CRUMP stars.
This is discussed in \S~\ref{S:optimum}.

%(group I)
Fig.~\ref{fiducial} also indicates that only a few CEMP stars are formed from the gas with $\feoh \gtrsim -3$. %3.2$.  
   In particular, it is unlikely that faint SNe contributed to the formation of the Group I CEMP stars with $\coh \gtrsim -1$, which demands the condensation of carbon in the ISM by more than 100 times than what is realized in our models. 
Although the majority of the Group I stars are made up of \cemps\ stars, which are generally thought to be formed by a binary mass transfer from low-mass AGB stars, there are more than ten CEMP-no stars with $\coh > -1$. 
Formation mechanism(s) other than faint SN is (are) required for these CEMP-no stars with very high carbon abundances. 
   We have argued in a separate paper that they are formed by a binary mass transfer in the same way as \cemps\ stars with differences in progenitor mass and the efficiency of the $s$-process nucleosynthesis in primary stars, where high-mass AGB stars and low efficiencies in the $s$-process nucleosynthesis as primary stars are required \citep[][see also Yamada et al. 2019b]{Komiya07}.

\begin{figure}
\includegraphics[width=\columnwidth,pagebox=cropbox]{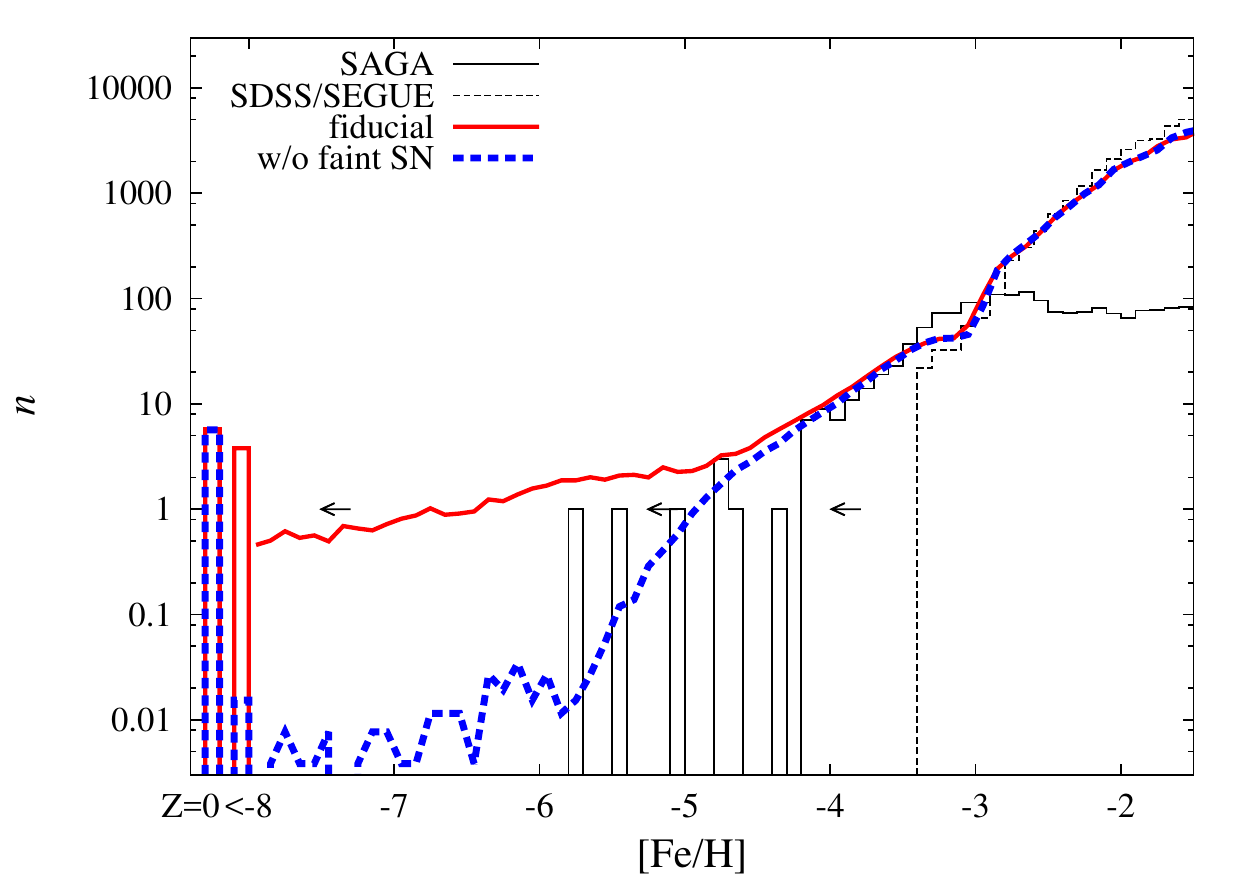}
\caption{
The metallicity distribution functions of low-mass stars that survive to date. 
Red and blue lines show the model results with and without faint SNe; 
two pillars on the left end denote the numbers of stars with $Z=0$ and with a finite metallicity of $\feoh < -8$, respectively. 
Solid and dashed histograms denote the observed MDFs for the SAGA database sample and for the SDSS/SEGUE data \citep{Carollo10}, respectively. 
The former sample stars have the accurate metallicity with the high resolution spectroscopy but are biased toward lower metallicity ($\feoh \lesssim -2.5$), while the latter is a homogeneous sample but has the metallicity only by medium resolution spectroscopy ($\feoh \gtrsim -3$). 
Horizontal arrows denote the upper limit of $\feoh$ for stars without the detection of iron. 
}\label{MDFfid}
\end{figure}

%(MDF)
In Fig.~\ref{MDFfid}, we compare the MDF of the low-mass survivors for the fiducial models with and without the faint SNe.  
Both model MDFs coincide with each other in the metallicity range of $\feoh \gtrsim -4$, and also predict a similar number of Pop III stars.  
The difference between these two model MDFs lies only in the number of HMP and lower metallicity stars below $Z \lesssim Z_{\rm cr}$, i.e., with $\feoh \lesssim -5$. 
The model with faint SNe produces an almost flat MDF at lower metallicity, while the model without the faint SNe produces an MDF that significantly decreases the number of HMP stars.
This is expected from the fact that smaller metallicity demands greater dilution of SN yields by the diffusion in the polluted regions.  
The difference in the number of HMP stars between the two models grows larger to be two orders of magnitude for the low metallicity of $\feoh <-6$. 
In the model without faint SNe, such low-metallicity stars are formed in the mini-halos, already polluted by the galactic wind, prior to the star formation (see below).  
In both MDFs, we see a small dent near $\feoh \simeq -3$, which is an artifact of averaging the chemical composition by the criterion of eq.~(\ref{eq:30mix}). 
Otherwise, the slope will be almost constant and a linearity relationship nearly holds between the metallicity and the number of low-mass survivors for $\feoh \gtrsim -4$. 

The observed MDFs in Fig.~\ref{MDFfid} represent the data taken from the SAGA database and the SDSS/SEGUE sample \citep{Carollo10} to compare with the model predictions. 
The SAGA database collects literature data of those stars which have the abundances derived from high- and medium-resolution spectra, which provides the largest sample of EMP stars.
It is to be noted that the determination of the metallicity of $\feoh \lesssim -3$ requires high-resolution follow-up observations.
   Since the high-resolution observations are biased toward the low metallicity of $\feoh \lesssim -2.5$, the sample stars are not complete at larger metallicities.  
On the other hand, the SDSS/SEGUE sample provides a relatively unbiased and homogeneous data.  
However, the metallicities of their sample stars are not reliable for $\feoh \lesssim -3$ since they are determined by the medium resolution spectroscopic data, and the sample size is small for $\feoh < -3$. 
It is argued that the SAGA database sample is almost unbiased below $\feoh \le -2.8$ \citep{Suda08}.
   Thus, we may scale the MDF from the SDSS/SEGUE sample to match the MDF from the SAGA database at the bin between $\feoh = -2.8$ to $-2.9$.
We combined two MDFs at $\feoh = -2.8$ below and above which the SAGA database and the SDSS/SEGUE is adopted, respectively. 
We scale them to match the model MDFs.  

%(\feoh > -4)
Both model MDFs with and without faint SNe are consistent with the observations at $\feoh \gtrsim -4$. 
%(UMP)
At $-5 < \feoh < -4.2$, both models tend to predict slightly more stars than observed. 
%(HMP)
For $\feoh < -5$, the model with faint SNe gives a much larger number of stars than observed by more than 1 dex.  
In contrast, the model without faint SNe achieves much better consistency with the observed MDF, though the observational sample of HMP stars is small,  
The most iron-poor star ever detected is SMSS J031300.36-670839.3 with $\feoh < -7.3$ \citep{keller14,bessel15}. 
In the model without faint SNe, the number of predicted stars below $\feoh = -7$ is very small, but it is expected to find $\sim 1$ star between $-\infty < \feoh < -7$. 

%(PopIII)
We note that $\sim 6$ Pop III stars are predicted regardless of the models with or without faint SNe. 
The absence of strict Pop III star in the current sample is indicative that the frequency of low-mass stars among Pop III stars is significantly lower than among HMP stars.
Although the shape of the IMF of Pop III stars is not well understood, higher mass IMFs are favored by theoretical studies.
As discussed in our previous studies and in later sections of this paper, the accretion of the interstellar gas polluted by SN yields after the birth can significantly change the MDF for $\feoh \lesssim -4$ and can explain the absence of Pop III stars and the lowest metallicity tail of the MDF. 

%%% \S 3.1.1 
\subsubsection{external enrichment of mini-halos}

The external enrichment has been discussed as a formation scenario for the most low-metallicity Pop II stars \citep{Smith15, Chiaki18a}. 
Energetic Pop III supernovae in mini-halos can blow away the metals from their host mini-halos, and enrich the IGM around the host mini-halos, and also, the nearby mini-halos with the metals. 
The nearby mini-halos or the mini-halos which accrete the metal-enriched IGM can form the second generation stars with very low metallicity. 

We found that the external enrichment plays a minor role in the fiducial model.
Only $1\%$ of HMP stars are the first-generation stars in the mini-halos enriched by the external-pollution. 
In our model, most of the ejected metals fall back to its progenitor proto-galaxy as the proto-galaxy increases its mass.

%%% \S 3.2
\subsection{Parameter Dependences}\label{S:param}

Our fiducial model fails to reproduce the distribution of stars on the $\coh \hyp \feoh$ plane. 
Here we try to find the parameter set that can reproduce the observations in the framework of the faint SN scenario.

%%% \S 3.2.1
\subsubsection{Mixing Mass of SN ejecta}\label{S:mix}

Initial carbon abundance, $\coh$, of the polluted region is in inverse proportion to the mixing mass as described in eq.~(\ref{eq:CoHup}). 
In models with smaller values of diffusion coefficient, $C_{\rm diff}$, and swept-mass, $ M_{\rm sw51}$, the second generation stars will have higher $\coh$.  
In this case, a larger number of Pop III stars are formed and explode as faint SNe since the progress of metal pollution in the ISM is slower than what is realized with larger values of $C_{\rm diff}$ and/or $ M_{\rm sw51}$.  

Figure~\ref{C_diff} shows the dependence on the diffusion coefficient, $C_{\rm diff}$.  %, M_{\rm sw}$ and $Z_{\rm cr}$. 
In the slow diffusion model with $C_{\rm diff} = 10^{-7} \msun \, {\rm yr}^{-3/2}$ (lower panel), the metals ejected by a SN are mixed and diffuse only over the mass of $\sim 10^{5}\msun$ after $10^8$ yrs. 
This model predicts higher $\coh$ for UMP/HMP stars than the fiducial model as a whole, and yet, most of them are still distributed below $\coh \lesssim -3 $.

%%% Fig. 7 
\begin{figure}
\includegraphics[width=\columnwidth,pagebox=cropbox]{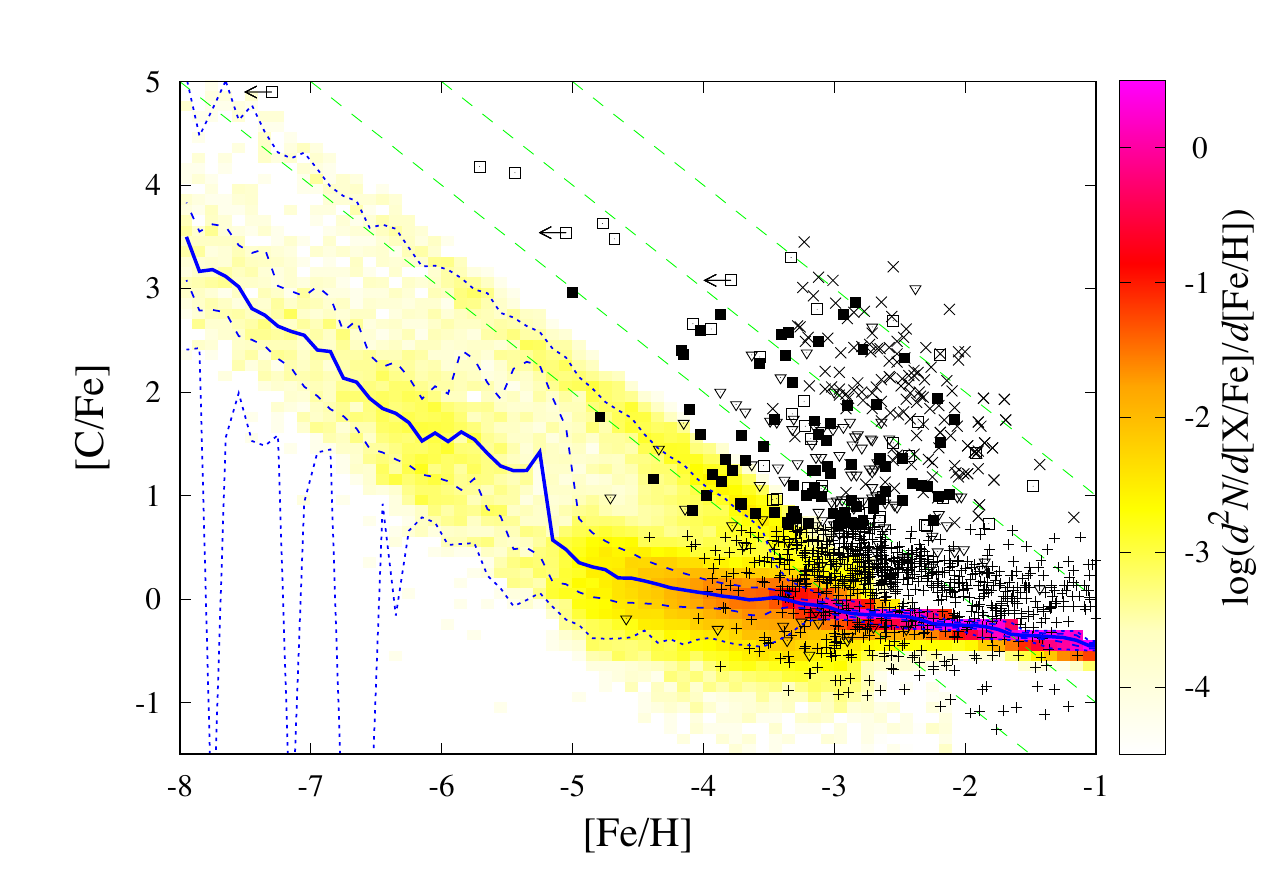}
\includegraphics[width=\columnwidth,pagebox=cropbox]{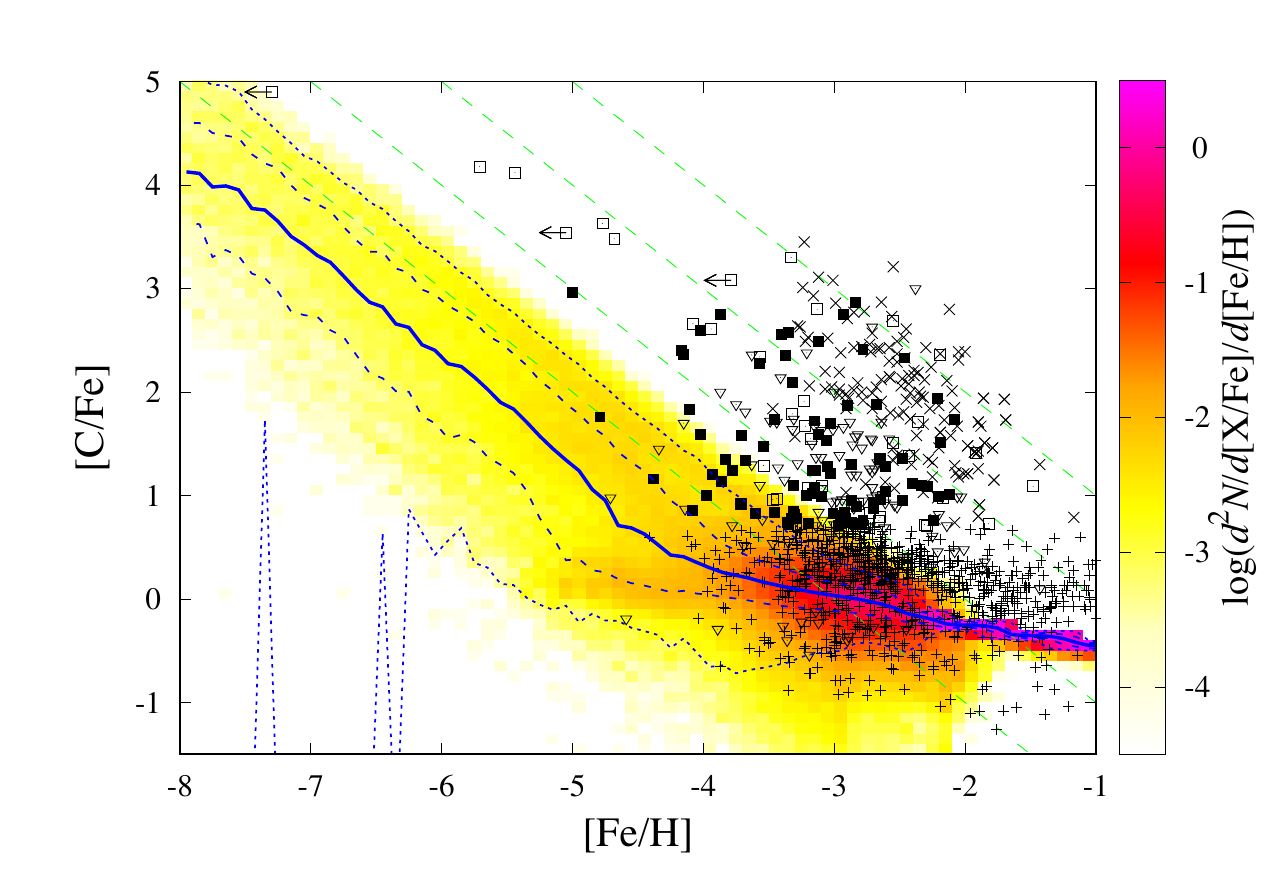}
\caption{
The same as Fig.~\ref{fiducial} but for the model of the diffusion coefficient, $C_{\rm diff} = 10^{-5} \, \msun \, {\rm yr}^{-3/2}$ (upper panel), and the model of $C_{\rm diff} = 10^{-7} \, \msun \, {\rm yr}^{-3/2}$ (lower panel).  
}\label{C_diff}
\end{figure}

%%% Fig. 8 
\begin{figure}
\includegraphics[width=\columnwidth,pagebox=cropbox]{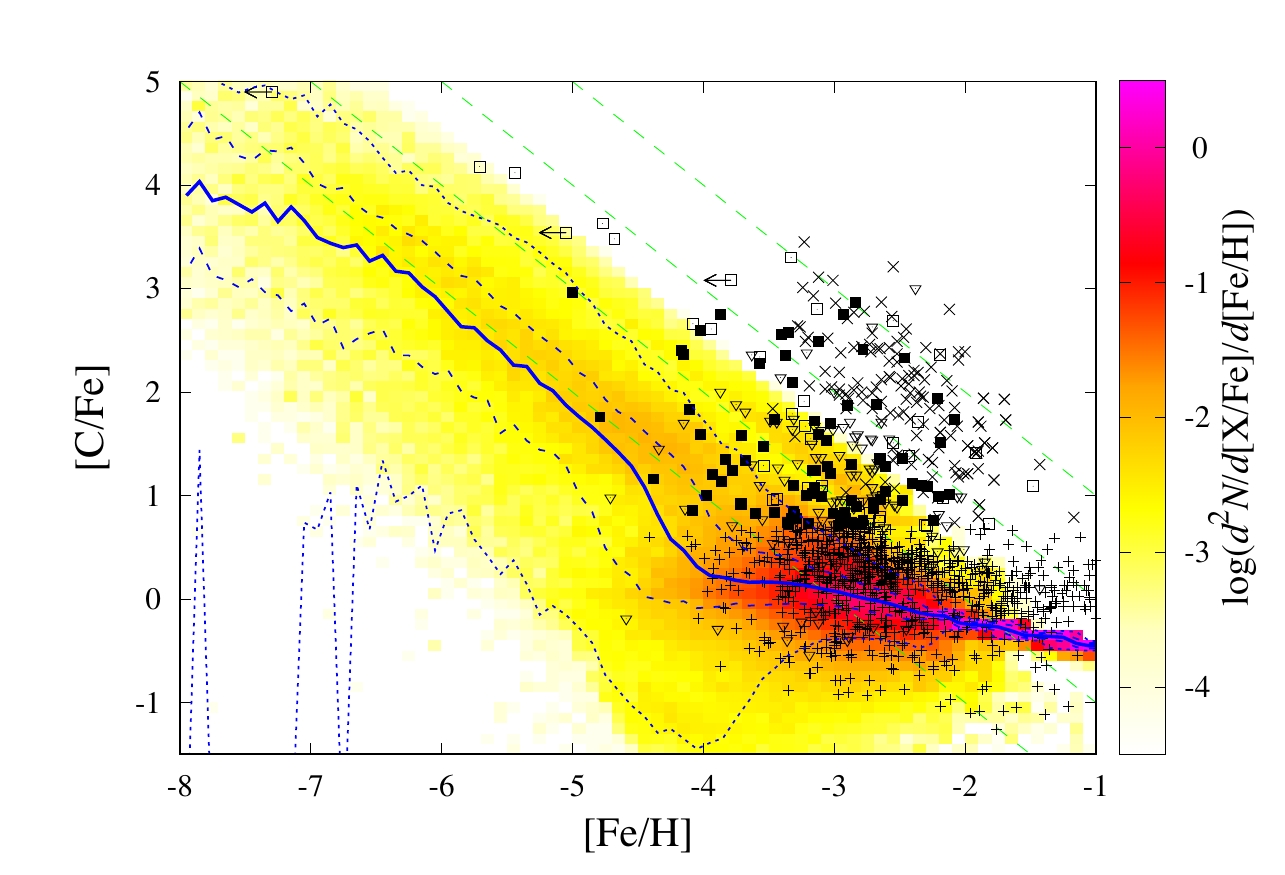}  
\includegraphics[width=\columnwidth,pagebox=cropbox]{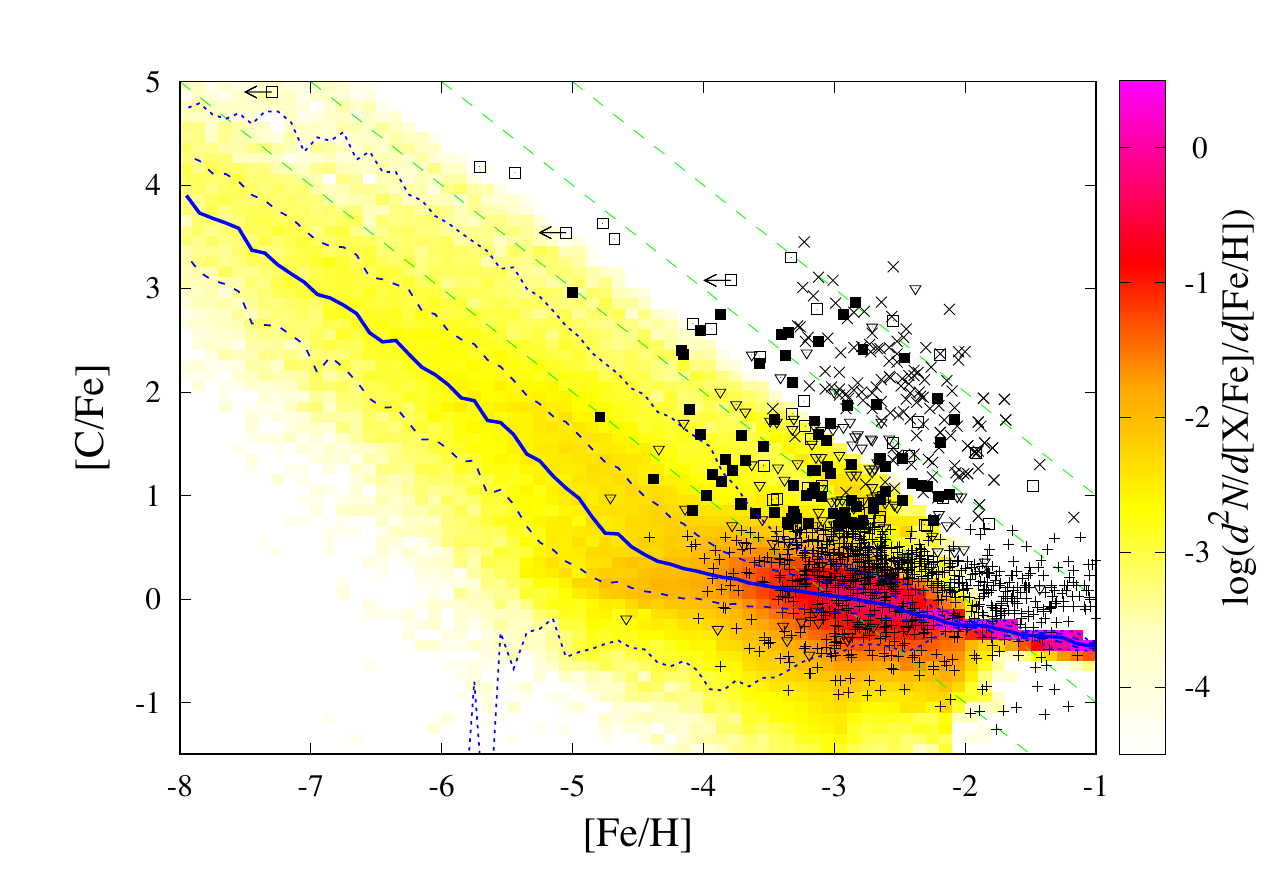}   
\caption{
The dependence on the swept-up mass, $M_{\rm sw51}$. 
In the top panel we adopt by a factor of 10 smaller mass, $M_{\rm sw51} = 5.1\times 10^3 \msun$. 
In the bottom panel, we assume the fiducial value of $M_{\rm sw51}$ but the explosion energy to be smaller by a factor of 10, i.e., $E_{\rm SN} = 10^{50} {\rm erg}$, for faint SNe. 
In the both models, the small diffusion coefficient, $C_{\rm diff} = 10^{-7} \msun \, {\rm yr}^{-3/2}$, is adopted. 
}\label{M_sw}
\end{figure}

%%% Fig. 9 
\begin{figure}
\includegraphics[width=\columnwidth,pagebox=cropbox]{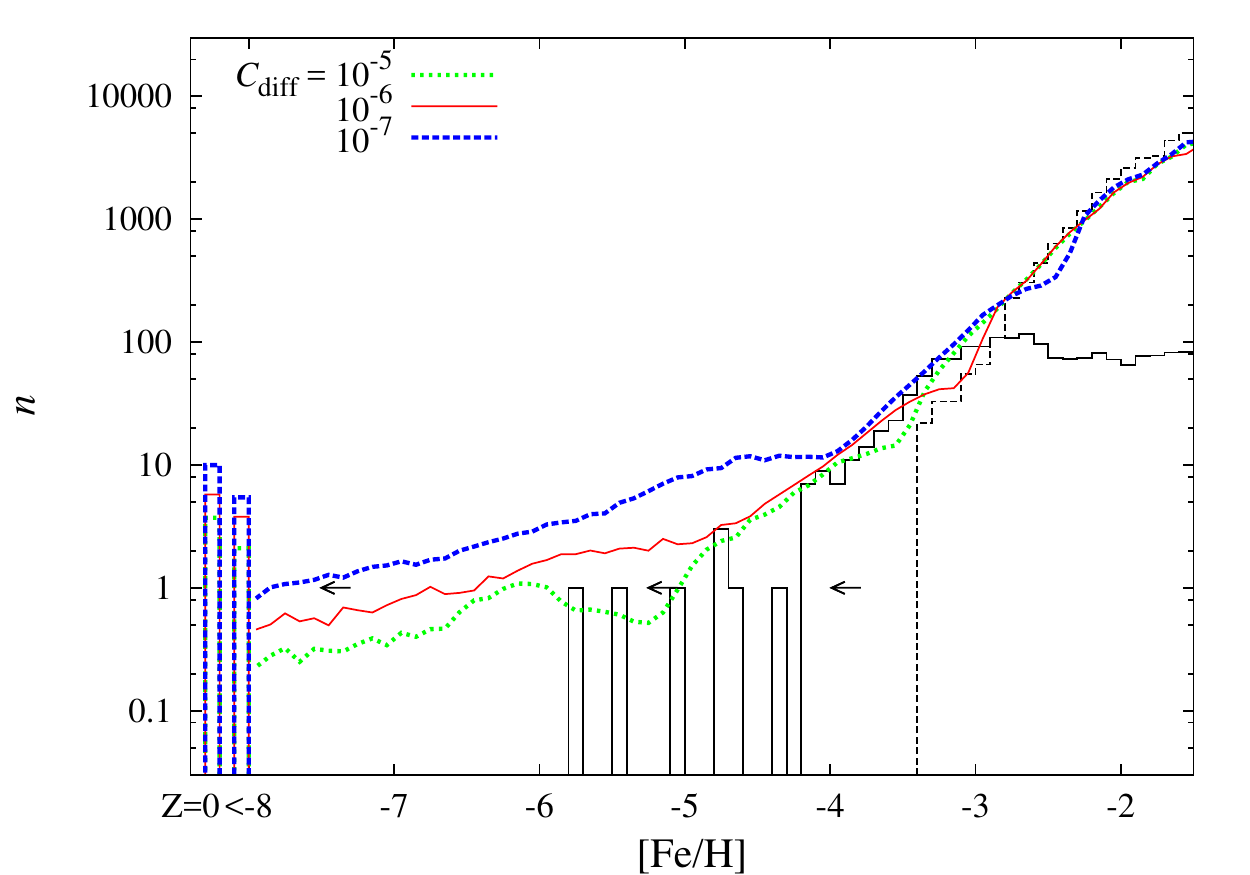}  
\includegraphics[width=\columnwidth,pagebox=cropbox]{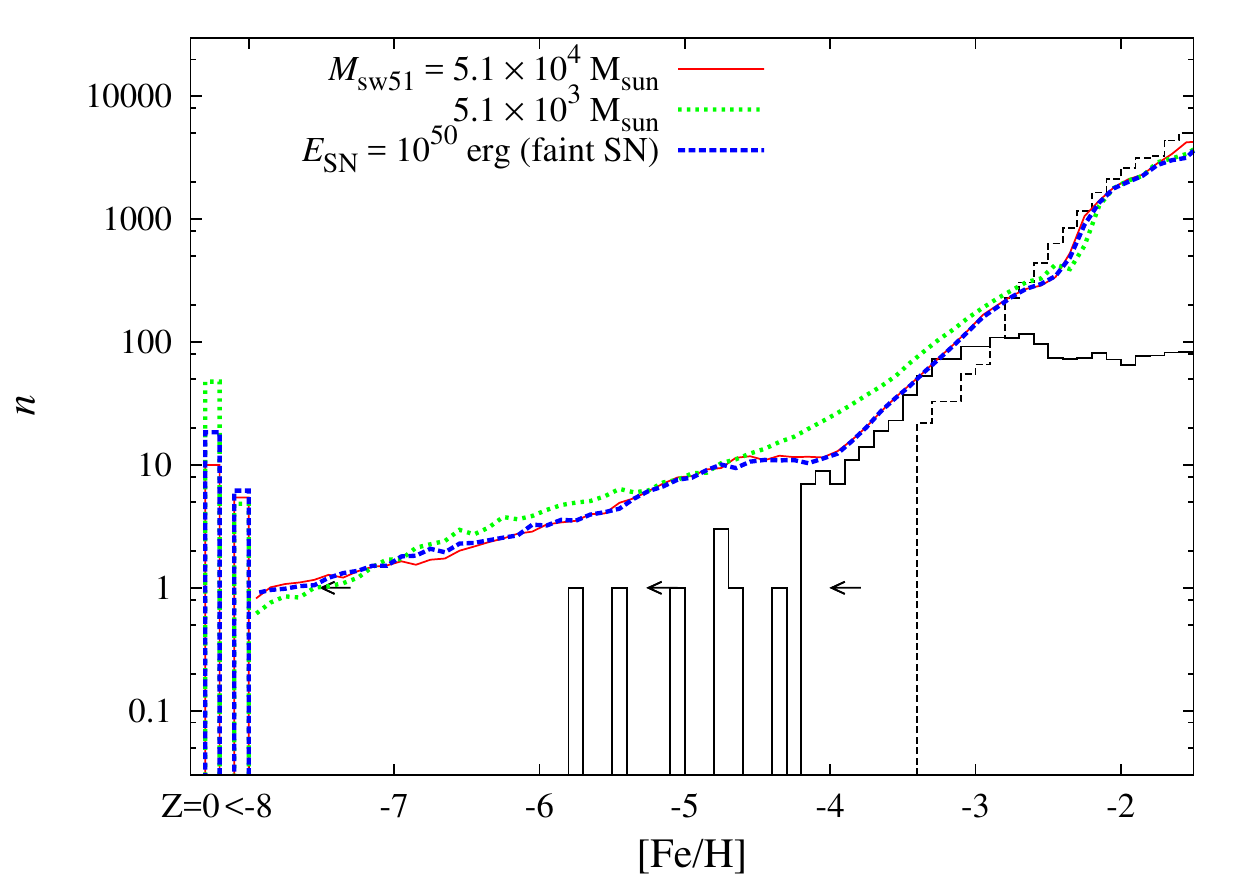}  
\caption{
The dependence of MDF on the $C_{\rm diff}$ (upper panel) and $M_{\rm sw}$ (lower panel). 
}\label{MDFmix}
\end{figure}

%(-2,5 branch)
In the fast diffusion model with $C_{\rm diff} = 10^{-5} \msun \, {\rm yr}^{-3/2}$ (upper panel), on the other hand, we see two branches for $\feoh < -4$.  
One is at slightly above $\coh \sim -3$ and the other is around $\coh \sim -4.5$. 
The former branch corresponds to the upper limit of $\coh$ given in eq.~(\ref{eq:CoHup}). %the carbon abundance of $\Delta M_{\rm C}/M_{\rm SNR}$. 
The majority of these stars are formed in the re-accreted gas, enriched by the outflow from its own proto-galaxy. 
A SN triggers the galactic wind from mini-halos and enriches the surrounding IGM, where the chemical composition of the wind is $\coh \sim -2.5$. 
Re-accreted gas onto the proto-galaxy contains metals contaminated by the outflow, driven by SNe from the former host galaxy.  
Since the metal diffusion is neglected for the re-accreted gas in the present computations, the stars formed of the re-accreted gas have carbon abundances similar to those of galactic wind. 
If we take into account the diffusion of re-accreted metals, the number of stars in the high $\coh$ branch will be smaller.
In the model with the small $C_{\rm diff}$, the lower branch is shifted toward higher $\coh$ and overlaps with the upper branch.

%(main branch)
The stars in the lower branch are formed in the proto-galaxy with gas mass of $M_{\rm gas} = 10^6$ - $10^7 \msun$.
The mixed ejecta of one or a few faint SN(e) form stars in this branch. 

%(E_SN)
In models with $ M_{\rm sw51} = 5.1 \times 10^{3}\msun$, stars with $\coh > -2$ can be formed, as shown in Figure~\ref{M_sw}. 
Since the upper limit of carbon abundances in eq.~(\ref{eq:CoHup}) increases to $\coh \simeq -1.5$, they have the comparable carbon abundance with the observed CRUMP stars. 
However, the number of stars with $ \coh > -2 $ is still very small, and vast majority of UMP/HMP stars have $ \coh < -3 $. 

The impact of reducing the explosion energy is similar to that of the lower $M_{\rm sw51}$ case as shown by the bottom panel of Fig.~\ref{M_sw}.
The combined effects of low-energy SNe and less-efficient mixing of ejecta with the ISM do not provide sufficient number of CRUMP stars with $\coh > -2$.
According to the models of \citet{Tominaga14}, most of the known CRUMP stars can be reproduced by the models with $E_{\rm SN} = 10^{51} {\rm erg}$.
These suggest that low-energy SNe are not necessarily responsible for the contribution to the CRUMP stars.

%(MDF, Cdiff)
Figure~\ref{MDFmix} shows the dependence of MDFs on $C_{\rm diff}$ (top panel) and the swept-up mass (bottom panel).  
The distribution at $\feoh > -4$ is independent of $C_{\rm diff}$ except for the dip due to the artificial averaging of chemical composition by the criterion of eq.~(\ref{eq:30mix}), which occurs at $\feoh \sim -2.5$, $-3$, and $-3.5$ for models of $C_{\rm diff} = 10^{-7}$, $10^{-6}$, and $10^{-5} \msun \, {\rm yr}^{-3/2}$, respectively.  
On the other hand, the number of UMP/HMP stars as well as Pop III stars is sensitive to $C_{\rm diff}$. 
In the case of smaller $C_{\rm diff}$, the metal enrichment process of proto-galaxies is delayed, which produces more faint SNe. 
As a result, more second generation UMP/HMP stars are formed. 
For $C_{\rm diff} = 10^{-8} \msun {\rm yr}^{-3/2}$, the result is quite similar to that for $C_{\rm diff} = 10^{-7} \msun {\rm yr}^{-3/2}$.  

%(MDF, Msw)
The smaller the value of $M_{\rm sw}$ (with $M_{\rm w}$ unchanged), the larger the number of Pop III stars since the ISM gas in proto-galaxy remains with $Z=0$. 
Smaller polluted mass around many faint SNe tend to produce metal-rich second generation stars due to the enrichment of ISM by other faint SNe.
In addition, most of the SN ejecta participate in outflows because the ratio, $M_{\rm w}/M_{\rm sw}$, becomes larger (see eq.~[\ref{eq:windmass}]).
The ejecta accrete onto their host mini-halos as they grow in mass, where we do not consider the metal diffusion, which preserve the region of $Z = 0$ and increase more chance for faint SNe in the host halo.
Both of these effects contribute to the increase in the number of stars around $\feoh \sim -4$. 
We found $\sim 1,600,000$ faint SNe in this case, which is $\sim 10$ times more than the case in the fiducial model.

%%% section \S 3.2.2 %%%
\subsubsection{Criterion for faint SN}\label{S:Zcr}

In the fiducial model, all the SNe are assumed to be faint SNe if the metallicity is below the critical metallicity, $Z_{\rm cr} = 10^{-5} \Zsun$.  
%(Zcr dependence)
Figures~\ref{Z_cr} and \ref{MDF_Zcr} show the results with $Z_{\rm cr} = 10^{-3} \Zsun$. 
We adopt the smaller diffusion rate of $C_{\rm diff} = 10^{-7} \msun \, {\rm yr}^{-3/2}$, and the smaller swept-up mass of $M_{\rm sw51} = 5.1 \times 10^3 \msun$ compared with the fiducial model in order to obtain more stars with larger $\coh$. 
As a result, more than a quarter of stars evolve to CEMP stars for $\feoh < -3.8$, and the majority of them have $\cfe > 0.7$ by the chemical enrichment up to $\feoh \simeq -4.2$. 
This result shows the possibility that the Group II stars with weak carbon enhancement are originated from faint SNe. 
However, as in the fiducial model, the predicted carbon abundances of CEMP stars are much lower than CEMP stars in the Group I or Group III.  
We need additional source of carbon in this framework.

%%% Fig. 10
\begin{figure}
\includegraphics[width=\columnwidth,pagebox=cropbox]{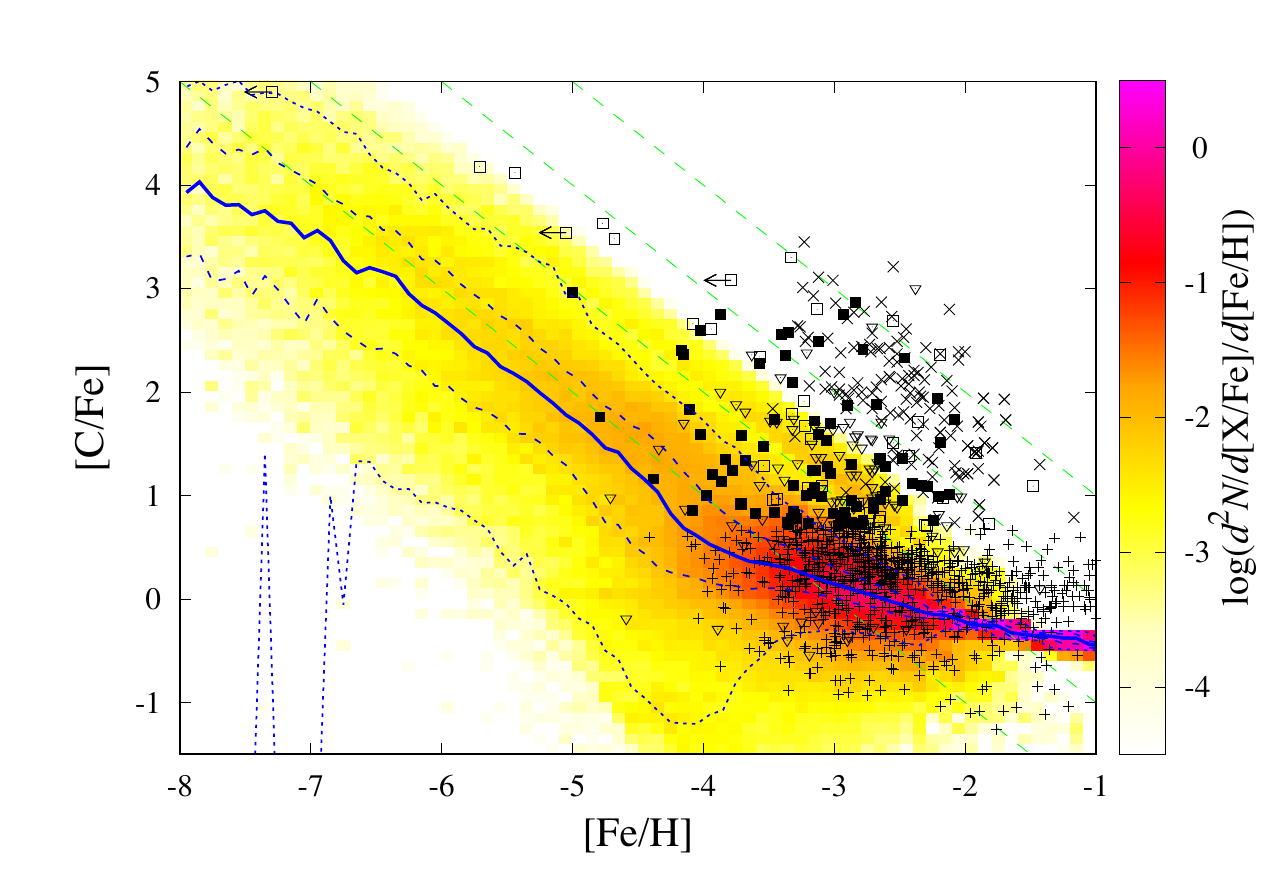}  
\caption{
The model with $Z_{\rm cr} = 10^{-3} \Zsun$. 
Other parameters are the same as in the case of the top panel of Fig.~\ref{M_sw} ($M_{\rm sw51} = 5.1 \times 10^3 \msun$ and $C_{\rm diff} = 10^{-7} \msun \, {\rm yr}^{-3/2}$). 
}\label{Z_cr}
\end{figure}

%%% Fig. 11
\begin{figure}
\includegraphics[width=\columnwidth,pagebox=cropbox]{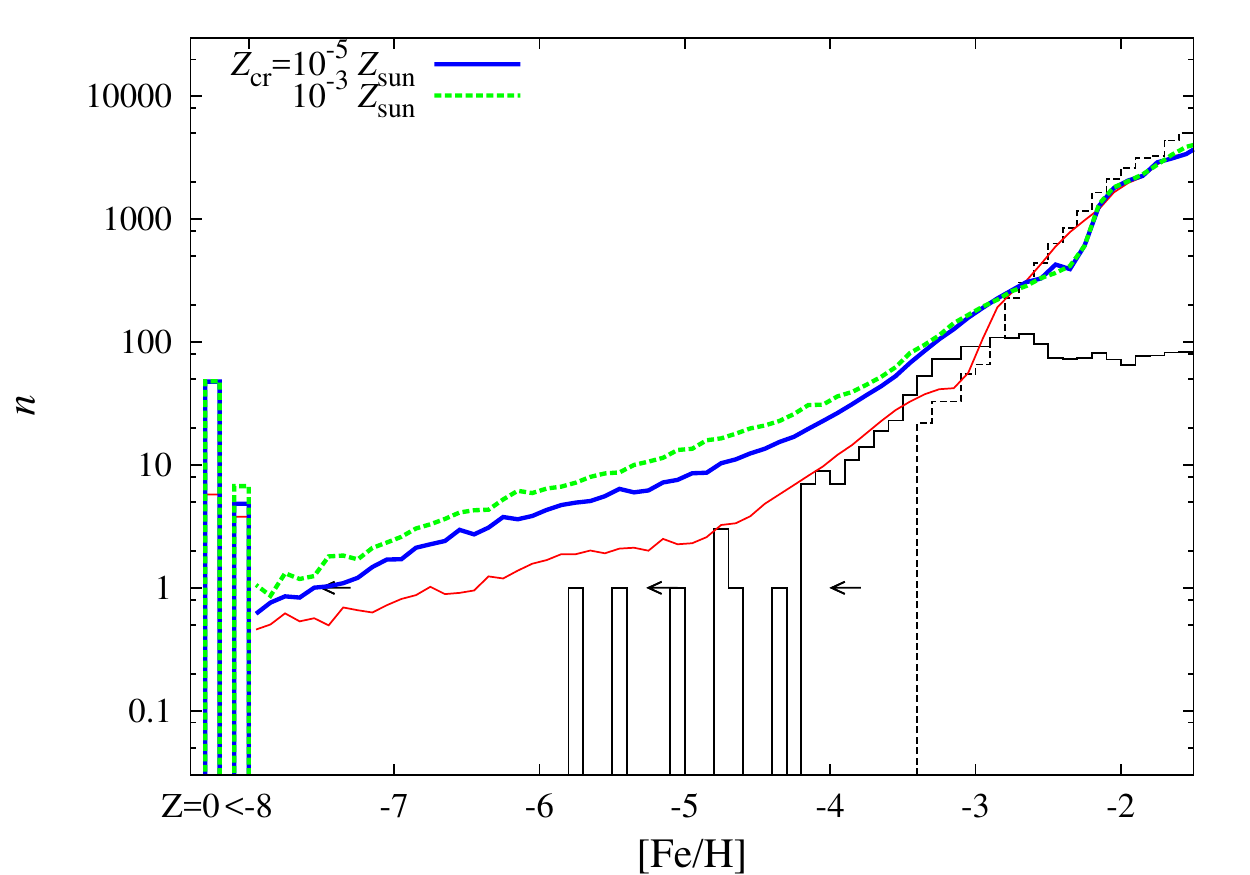}  
\caption{
The metallicity distribution functions of the model with $Z_{\rm cr} = 10^{-3}$ (green) and $Z_{\rm cr} = 10^{-5}$ (blue), for which small mixing parameters are assumed ($C_{\rm diff} = 10^{-7} \msun {\rm yr}^{-3/2}$ and $M_{\rm sw} = 5.1 \times 10^3 \msun$). 
We also plot the result of the fiducial model (red).  
}\label{MDF_Zcr}
\end{figure}

The metallicity of the second generation stars is controlled by the fallback parameter $f_{\rm fb}$ (defined in \S~\ref{S:faintResult}) for which all the Pop III stars end their lives as faint SNe with log-flat distribution in the metallicity range of $\feoh = -5$ and $-1$.  
%(f dependence)
Changing the distribution of $f_{\rm fb}$ does not solve the discrepancy of the $\coh$ values between the model and observed stars because $f_{\rm fb}$ only changes iron abundances, keeping carbon abundances unchanged.
Furthermore, we should note that the assumption on $f_{\rm fb}$ provides an optimistic case for the number of CEMP stars because the number of carbon-normal stars will increase by considering the case that Pop III stars end their lives as normal SNe.

Theoretically, the distribution of $f_{\rm fb}$ is a determinant of the number ratio of HMP/UMP stars, provided that most of them are descendants of first generation stars ending their lives by faint SNe.
The larger the value of $f_{\rm fb}$, the smaller the metallicity of a second generation star, and hence increases the number of HMP stars relative to UMP stars.
Observationally, the number of CEMP stars with $\cfe \sim 1$ is more abundant than CEMP stars with very high $\cfe$ of $3$ to $4$ \citep{Suda11,Norris13}. 
This indicates that a hihger value for $f_{\rm fb}$ is preferred.

%%% \S 3.2.3
\subsubsection{Pop III IMF}\label{S:IMF}

%%% Fig. 12
\begin{figure}
\includegraphics[width=\columnwidth,pagebox=cropbox]{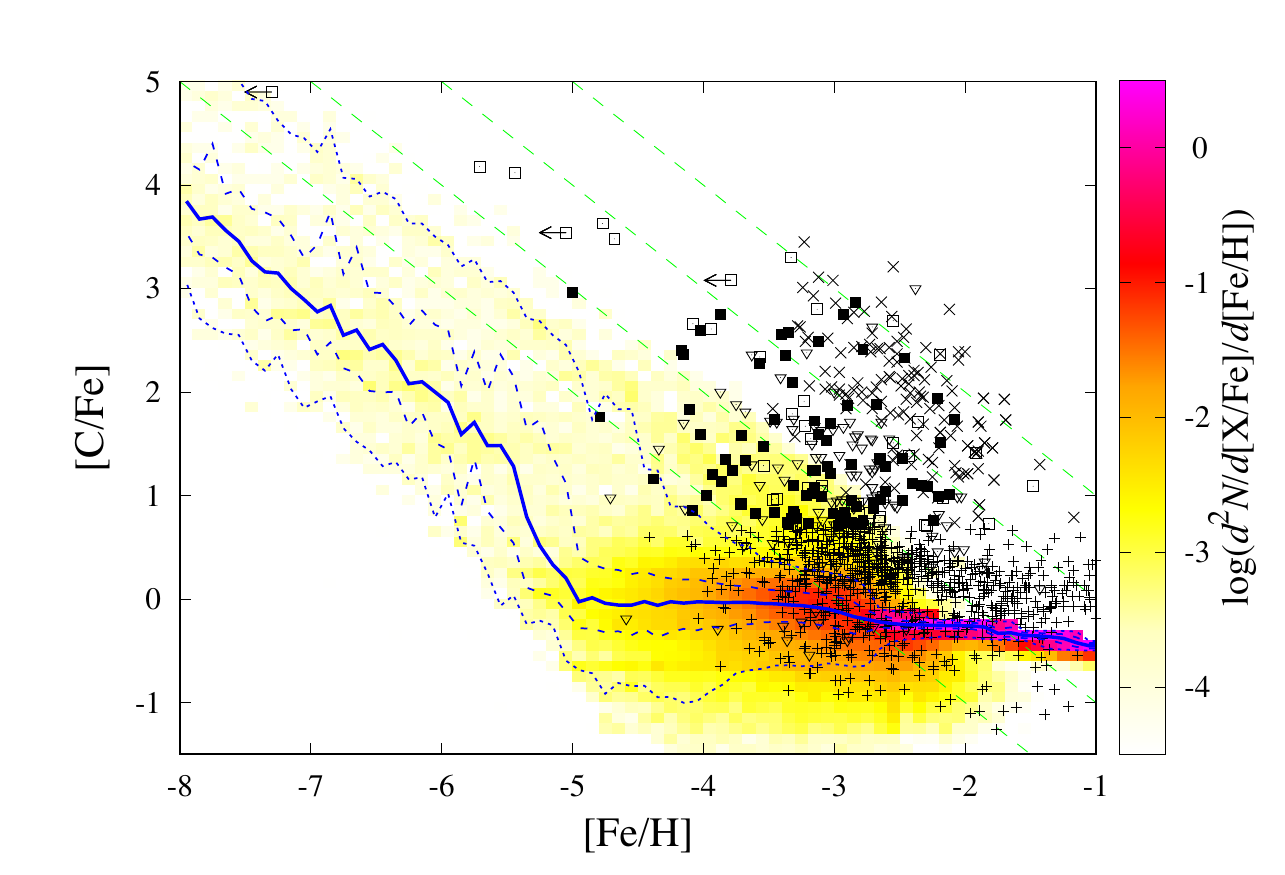}
\caption{
The same as Fig.~\ref{fiducial} but for the model with the IMF of the peak mass $\mmd = 100 \msun$ for Pop III stars. 
}\label{IMF100}
\end{figure}

%%% Fig. 13
\begin{figure}
\includegraphics[width=\columnwidth,pagebox=cropbox]{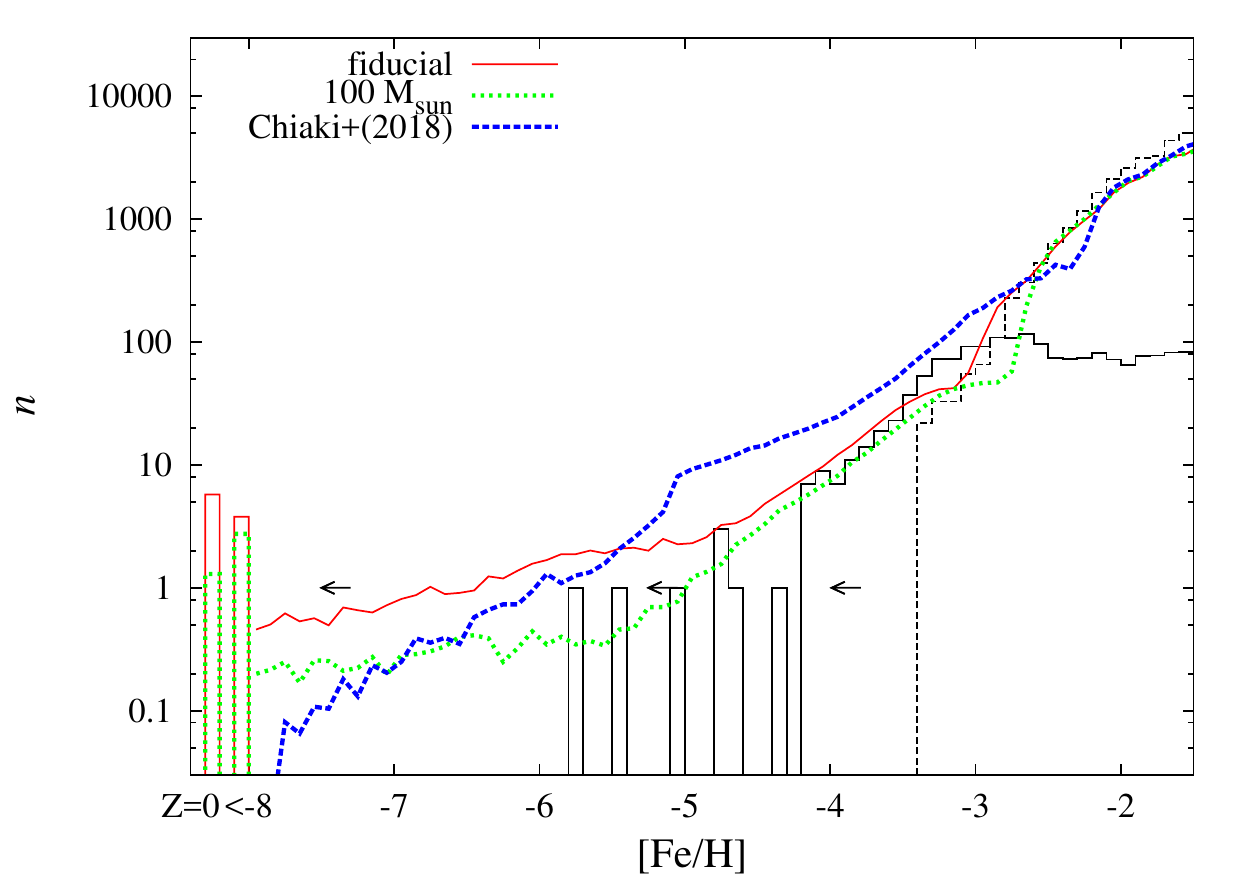}
\caption{
The metallicity distribution functions of the model with $M_{\rm md,Pop3} = 100 \msun$, where $C_{\rm diff} = 10^{-7} \msun \, {\rm yr}^{-3/2}$ $M_{\rm sw51} = 5.1 \times 10^{3} \msun$, and $Z_{\rm cr} = 10^{-4} \Zsun$ are adopted (green line). The model with the low-mass star formation criterion of \citet{Chiaki18b} (blue line) are compared with that of the fiducial model (red line). 
}\label{MDFpop3}
\end{figure}

%(Mmd=100)
Changing the IMF has an impact on the values of \cfe\ through the contribution from progenitors with different initial masses.
Since the carbon yield is not sensitive to the progenitor mass, 
carbon abundances in HMP/UMP stars are almost independent of the initial mass function in considering core-collapse SNe. 
%(PISN) 
In contrast, pair-instability SNe (PISNe) can be important if we consider the IMF including very massive stars.  
Figures~\ref{IMF100} and \ref{MDFpop3} show the carbon enhancement and the MDF for the model with the IMF of $\mmd = 100\msun$ for Pop~III stars.  
We assume that massive stars with $50 \hyp 140 \msun$ fail to explode and collapse to black holes, and stars with $140 \hyp 270 \msun$ explode as PISNe. 
In this model, $\sim 14,000$ PISNe take place. 
The number of CEMP stars and lower-metallicity stars including HMP/UMP stars are relatively smaller than that in the fiducial model.
This is partly because the frequency of faint SNe decreases.
Another factor is the competition between the large production of iron from single PISN and the pollution of large area with metals by the ejecta of a PISN. 

It is argued that a PISN of $\sim 150 \msun$ ejects larger amount of carbon but smaller amount of iron compared with CCSNe \citep{Umeda02}. 
These PISNe will contribute to the stars with the highest carbon abundance of $\coh \sim -2$. 
However, this is not supported by the abundance patterns of other elements in CEMP stars. 
The PISN yields show large enhancement of elements with even atomic numbers from carbon through calcium, and deficiency of odd $Z$ elements, which is not the case for CEMP stars. 

Changing the IMF for EMP stars does not have a significant effect.
The change of the mean peak mass for EMP stars is not subject to a great uncertainty because the supernova yields are not so sensitive to the initial mass, as stated above.
For instance, the mean peak mass of $10 \msun$ does not change the result very much.
On the other hand, the peak mass of $40 \msun$ may enhance the fraction of C-rich stars.
However, this is not a favored assumption from the constraints on the number of known EMP stars at present, which is discussed in \citet{Komiya09a}.

%%% \S 3.2.4
\subsubsection{Star Formation Efficiency}

%%% Fig. 14
\begin{figure}
\includegraphics[width=\columnwidth,pagebox=cropbox]{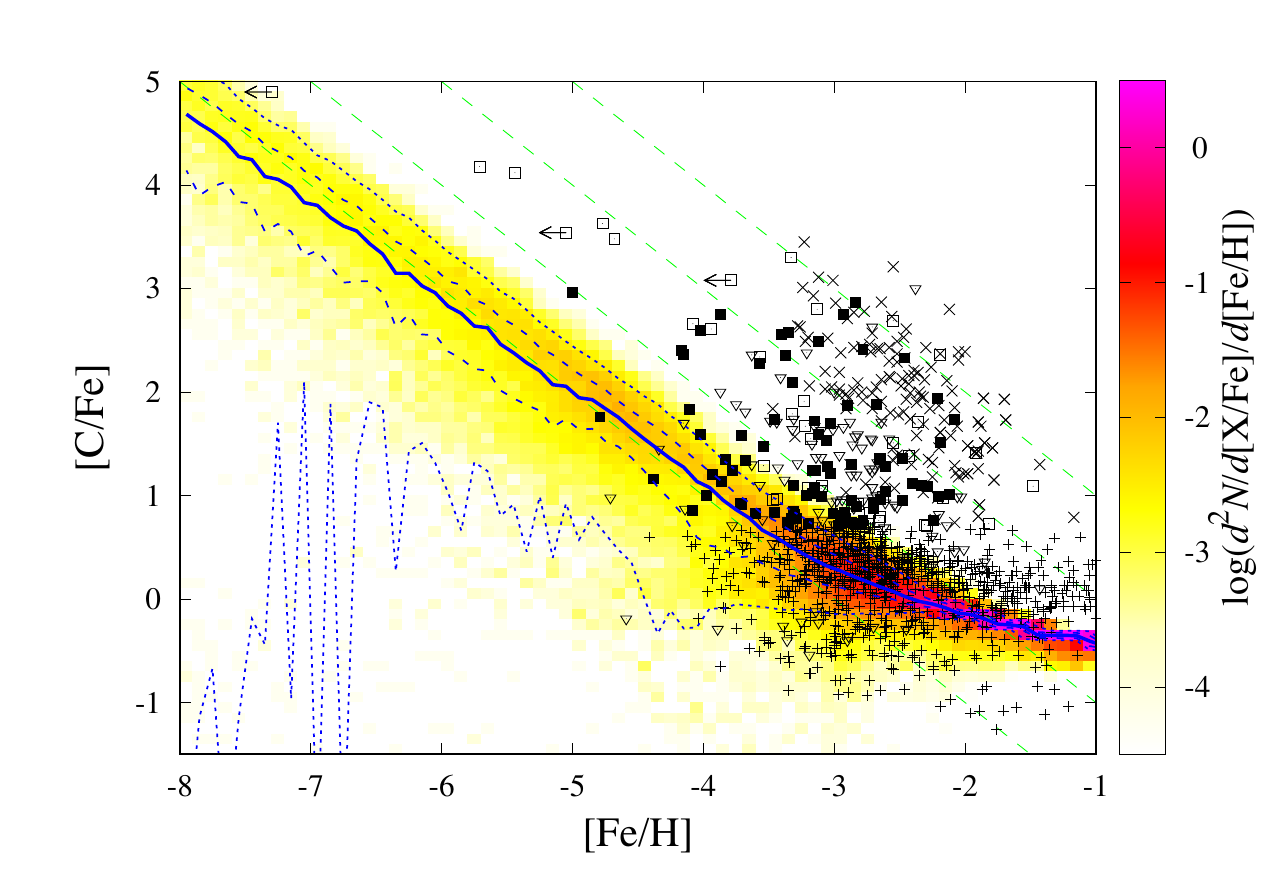}
\includegraphics[width=\columnwidth,pagebox=cropbox]{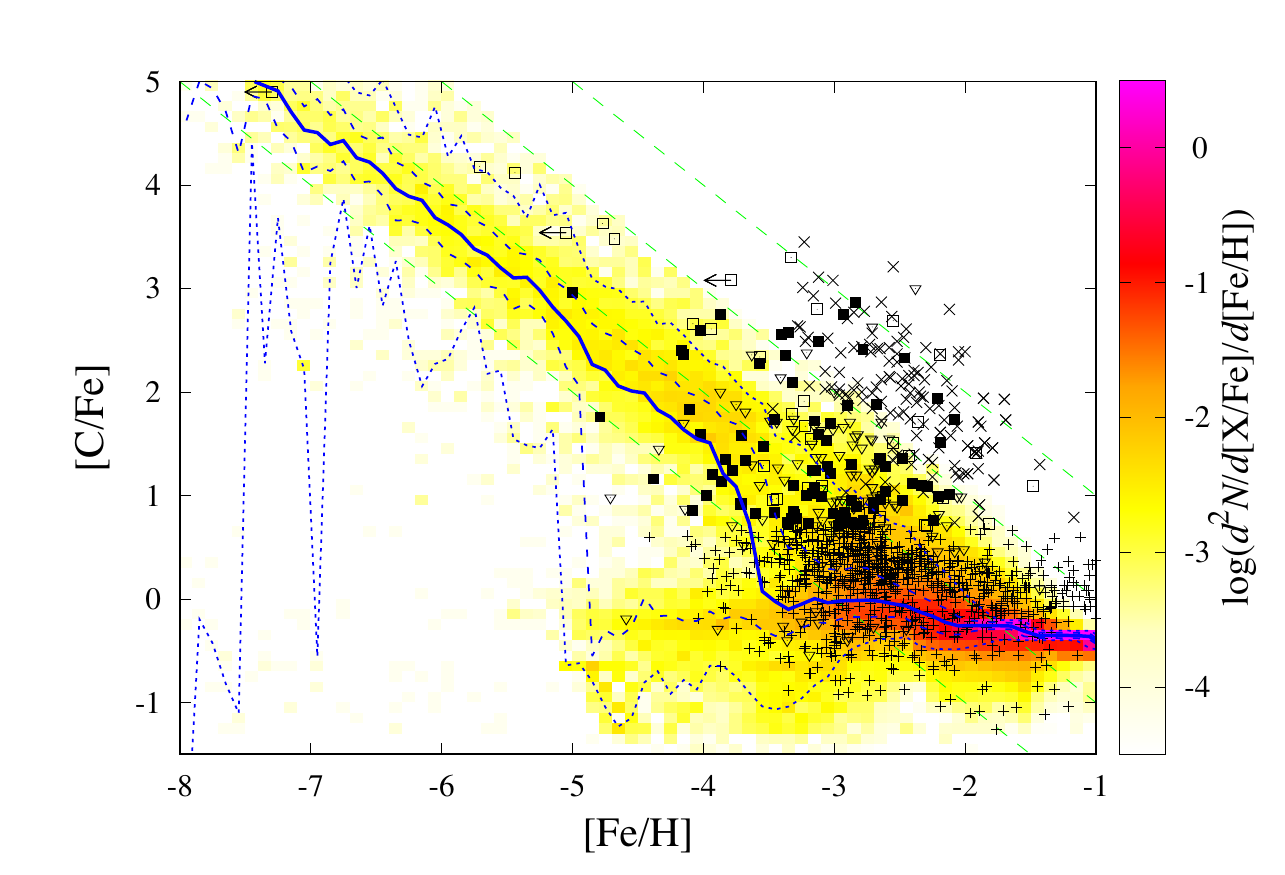}
\caption{
Models with a constant SFE ($\epsilon_{\star} = 5 \times 10^{11} {\rm yr}^{-1}$). 
Other parameters are fixed to the fiducial values (top panel), and the model with $M_{\rm sw51} = 5\times 10^3 \msun$, $Z_{\rm cr}=10^{-4}\Zsun$ and $M_{\rm md} = 50 \msun$ for the Pop~III stars (bottom panel). 
}\label{SFE}
\end{figure}

%%% Fig. 15
\begin{figure}
\includegraphics[width=\columnwidth,pagebox=cropbox]{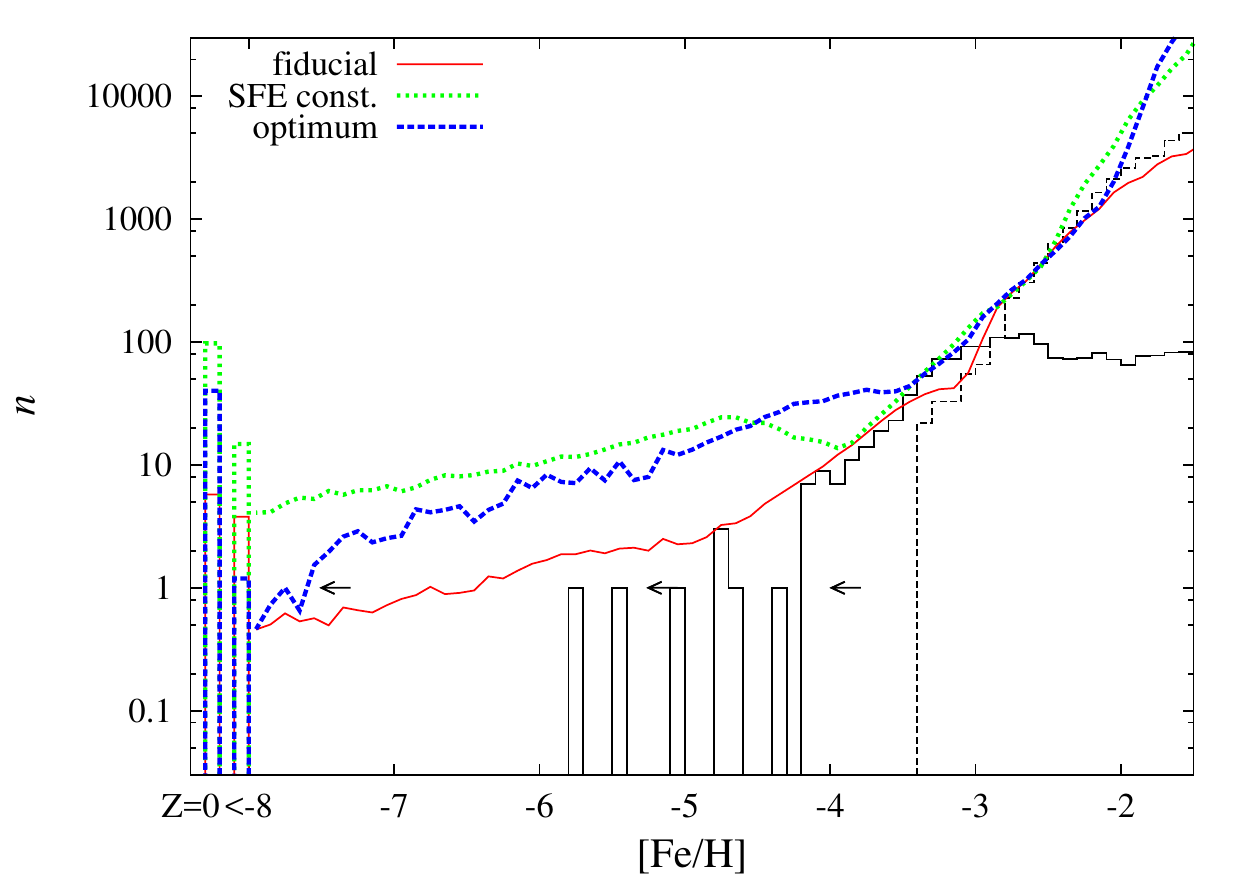}
\caption{
MDFs of constant SFE models. 
The green and blue lines represent the models in the top and bottom panels of Fig.~\ref{SFE}, respectively. 
  }\label{MDF_SFE}
\end{figure}

A constant Star Formation Efficiency (SFE) is examined in this subsection.
We adopt mass dependent SFE in the fiducial model, i.e., the star formation rate per unit gas mass equals to $\epsilon_{\star} \propto M_{\rm h}^{0.3}$ following \citet{Komiya16}. 
Lower values of SFE are favored in low-mass galaxies to reproduce the mass-metallicity relation of dwarf galaxies, and to explain the origin of $r$-process elements in EMP stars through nucleosynthesis in merging neutron stars \citep{Komiya16}. 
On the other hand, the SFE in proto-galaxies at high redshift is not yet well understood and is parametrized in this study. 

We show the model results with a constant SFE in Figures~\ref{SFE} and \ref{MDF_SFE}.  
The value of SFE is chosen to yield the solar metallicity at $z=0$ both for the constant and the mass dependent SFE, which gives $\epsilon_{\star} = 5 \times 10^{-11} yr^{-1}$ and $1.2 \times 10^{-14} ({M_{\rm h}/}{\msun})^{0.3}$, respectively. 
The star formation rate for constant SFE is higher than that for the mass-dependent SFE by $\sim 2$ dex in mini-halos of mass $M_{\rm h}=10^6 \msun$.
The higher star formation rate produces more stars before the metal diffusion becomes efficient in low-mass mini halos.
The change of SFE helps to increase carbon abundances by increasing the number of faint SNe, but it is still difficult to form stars with high $\coh$ with these models.
The distribution of HMP/UMP stars is concentrated around $\coh \sim -3$ by employing the fiducial values for other parameters. 

Also, more SNe in the low-mass mini-halos result in more efficient gas outflow. 
It reduces the gas mass in proto-galaxies, and hence, the number of EMP stars with $\feoh \sim -3$.
The number of such EMP stars is 0.6 times smaller in the constant SFE model than in the fiducial model. 
Therefore, the number of HMP/UMP stars is apparently large in the case of constant SFE because the MDFs are scaled to match the observations at $\feoh \sim -3$ in Fig.~\ref{MDF_SFE}.

%%% \S 3.2.5
\subsubsection{Optimum models}\label{S:optimum}

From the parameter dependences discussed above, we may choose the best fit parameters to reproduce the observed distribution of carbon abundances of CEMP stars.  
In the bottom panel of Fig.\ref{SFE}, we present the result of the model with the constant SFE, the small swept-up mass ($M_{\rm sw51}=5.1 \times 10^3 \msun$), the large critical metallicity for faint SNe ($Z_{\rm cr} = 10^{-4}\Zsun$, and the massive IMF ($\mmd = 50\msun$).   
The predicted distribution of CEMP stars almost covers the observations except for the Group I CEMP stars with $\coh \gtrsim -1$.  
The difficulty of the model is the overproduction of HMP/UMP stars due to the small diffusion rate.  
The fraction of HMP/UMP stars can be decreased by large diffusion coefficients but it reduces the overall carbon abundances.  
In this model, most of HMP stars show large carbon enhancement with $\coh > -3$.
In particular, the most carbon-enhanced stars are produced by PISNe, as mentioned in Sec.~\ref{S:IMF}.  
However, this result is likely to be a numerical artifact.
The main reason for the large carbon enhancement in PISNe is that we ignored the mixing of ejecta by diffusion process for infall-gas, as described in Sec.~\ref{S:mix}.
Accordingly, the gas with large \coh\ is localized in mini-halos that experienced PISNe.

Another way to produce carbon enhancement in HMP stars is to allow the formation of low-mass stars by setting a critical carbon abundance. 
This is the case by \citet{Chiaki18b} who gave the critical carbon and iron abundances, based on the observed abundance distribution of EMP stars, and on the gas cooling by carbon grains and silicate grains.  
They proposed that low-mass stars are formed only when  
\begin{equation}
10^{\coh - 2.30} + 10^{\feoh} > 10^{-5.07}. 
\end{equation}

%%% Fig. 16
\begin{figure}
\includegraphics[width=\columnwidth,pagebox=cropbox]{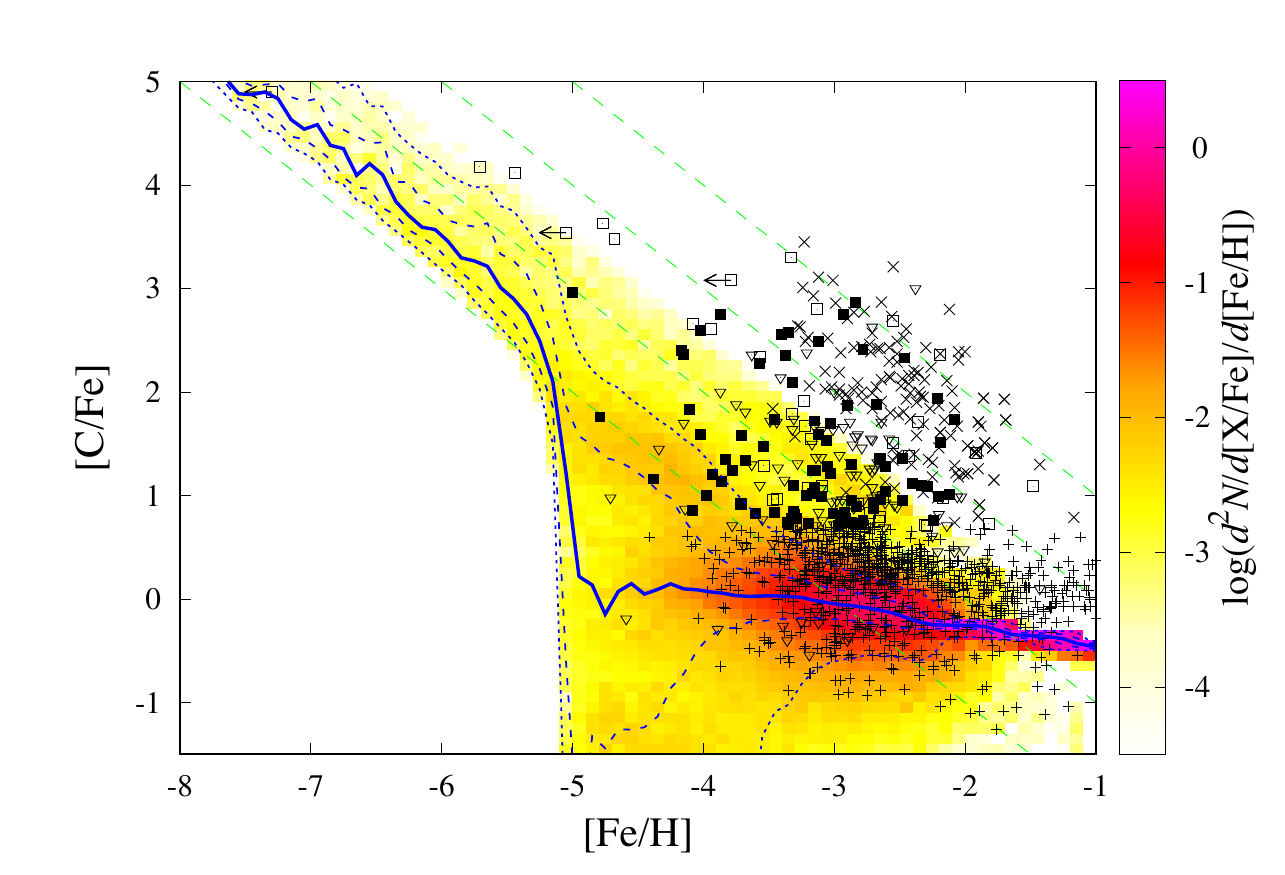}
\caption{
The model with low mass star formation only for $10^{\coh-2.30}+10^{\feoh} > 10^{-5.07}$ \citep{Chiaki18b}. 
Other parameters are chosen to favorably form CRUMP stars (see text).  
  }\label{optimum}
\end{figure}

We adopt this criterion on the model with the other parameters set favorable to form CRUMP stars; 
 $Z_{\rm cr} = 10^{-4} \Zsun$, 
 $C_{\rm diff} = 10^{-7} \msun \, {\rm yr}^{-3/2}$, and 
 $ E_{\rm SN} = 10^{50} {\rm erg}$ for faint SNe.   
The result is shown in Figure~\ref{optimum}. 

%(Chiaki+18 result)
In this model, HMP stars are distributed above $\coh > -3$ as a natural consequence of the criterion, but the distribution of stars with the carbon and iron abundances above the criterion is almost identical to the model assuming the low-mass IMF for Pop III stars. 
We see a discrepancy between the model result and the observations. 
At $-5.07 < \feoh < -4$, the model overpredicts the number of low-mass stars, and the number of stars drops below $\feoh < -5.07$, as shown in Figure~\ref{MDFpop3}. 
In addition, the predicted median of $\cfe$ is $\sim 0$ in this metallicity range.  

%(Discussion)
Our results do not necessarily reject the possibility of fitting the abundance distributions of HMP/UMP stars in the framework of the faint SN scenario.  
It is possible that the IMF is dependent on both the carbon and iron abundances, and we can construct such an abundance-dependent IMF that can reproduce the observed abundance distribution. 
However it requires an extremely inefficient metal-mixing in combination with the fine tuning of the low-mass star formation rate for HMP/UMP stars. 
Even after accepting the difficulty, CEMP-no stars in the Group I are still difficult to reproduce. 
A possible way to support the faint SN scenario will be an SN-triggered star formation \citep{Tsujimoto99}.
If the second-generation low-mass stars are formed in the gas shell swept up by the ejecta of faint SNe, both the abundances and frequency of HMP/UMP stars may be reproduced, although we do not have a reliable model for the SN-triggered star formation.
There are no predictions for the IMF of the second-generation stars in this model as well as the possibility of this star formation channel.
These theoretical uncertainties involve the argument of the definition and the choice of model parameters, and may encounter the same problem of the fine tuning of free parameters.
In addition, we need to explain the scarcity of \cemps\ stars for $\feoh < -3.5$, the origin of which is attributed to the mass transfer from AGB stars to low-mass stars in the binary systems.

%%% \S 4
\section{Binary Scenario}\label{S:binary}

As an alternative scenario for the CEMP-no stars, we examine the binary scenario in which a mass transfer from AGB stars is responsible \citep{Suda04, Komiya07}. 

We adopt the same assumptions for binary parameters as in \citet{Komiya07}.  % at $\abra{Z}{H} \geq -3$. 
We apply the Bondi-Hoyle accretion for the stellar wind, which blows with the velocity of 20 km ${\rm s}^{-1}$ from AGB stars whose surface carbon abundance have been enhanced to $\coh = 0$. 
The accreted matter is mixed in the surface convection zone with the depth of $0.2 \msun$. 

The log-normal period distribution function is adopted, where the peak of the distribution is at $P = 10^{4.8}$ days with the dispersion of $\sigma_P = 2.3$, which is applicable to low-mass stars in the solar neighborhood \citep{DM91}. 
%(CEMP-s/no)
Observationally, however, it is known that the fraction of \cemps\ stars drops below $\feoh \sim -3$. 
This is indicative that the binary properties of lower metallicity stars are different from more metal-rich stars, since the progress of $s$-process nucleosynthesis depends less on the metallicity for $\feoh \lesssim -2$ \citep{Yamada19a}. 
%(UMP)
We treat $P$ and $\sigma_P$ as free parameters and search the values which can reproduce the observations.

%%% Fig. 17 
\begin{figure}[htp]
\includegraphics[width=\columnwidth,pagebox=cropbox]{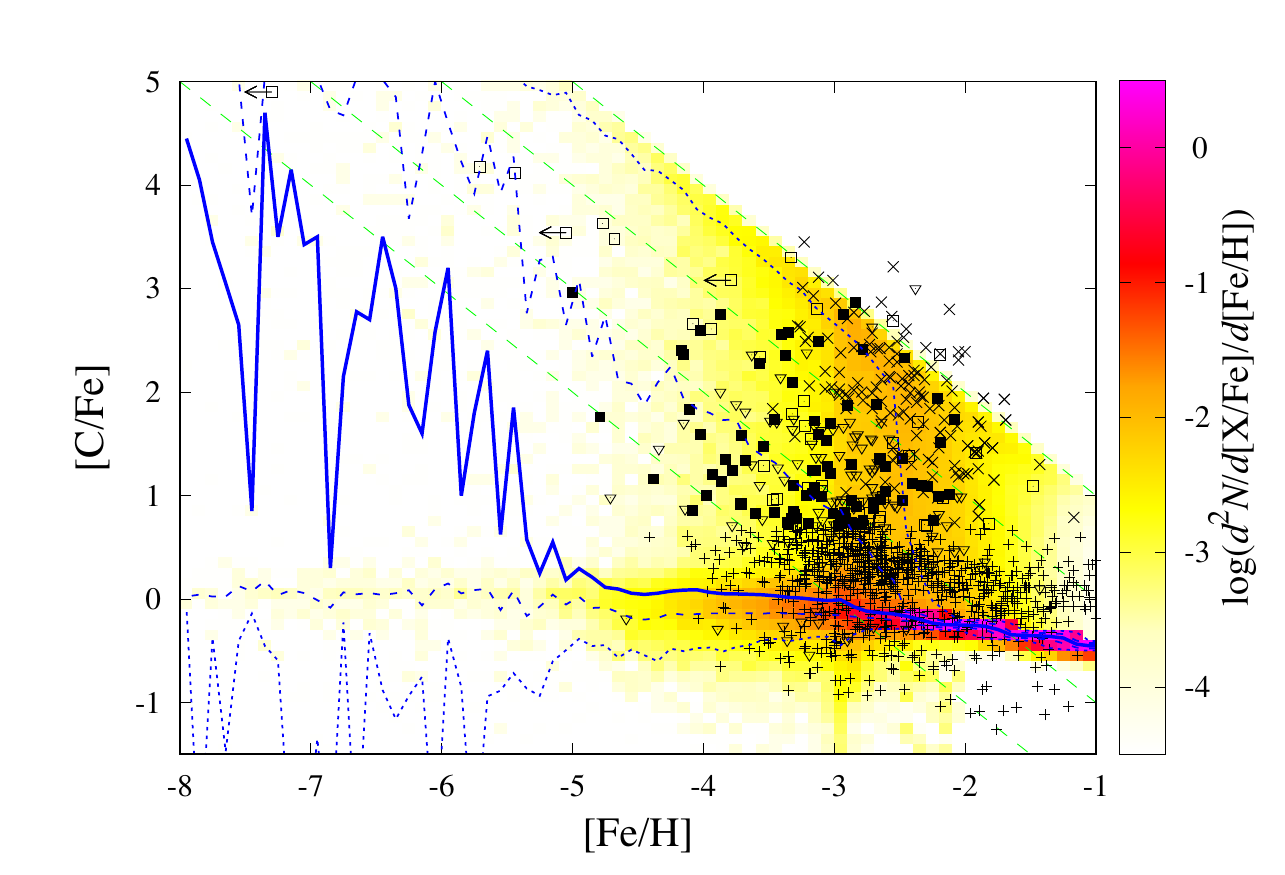}
\includegraphics[width=\columnwidth,pagebox=cropbox]{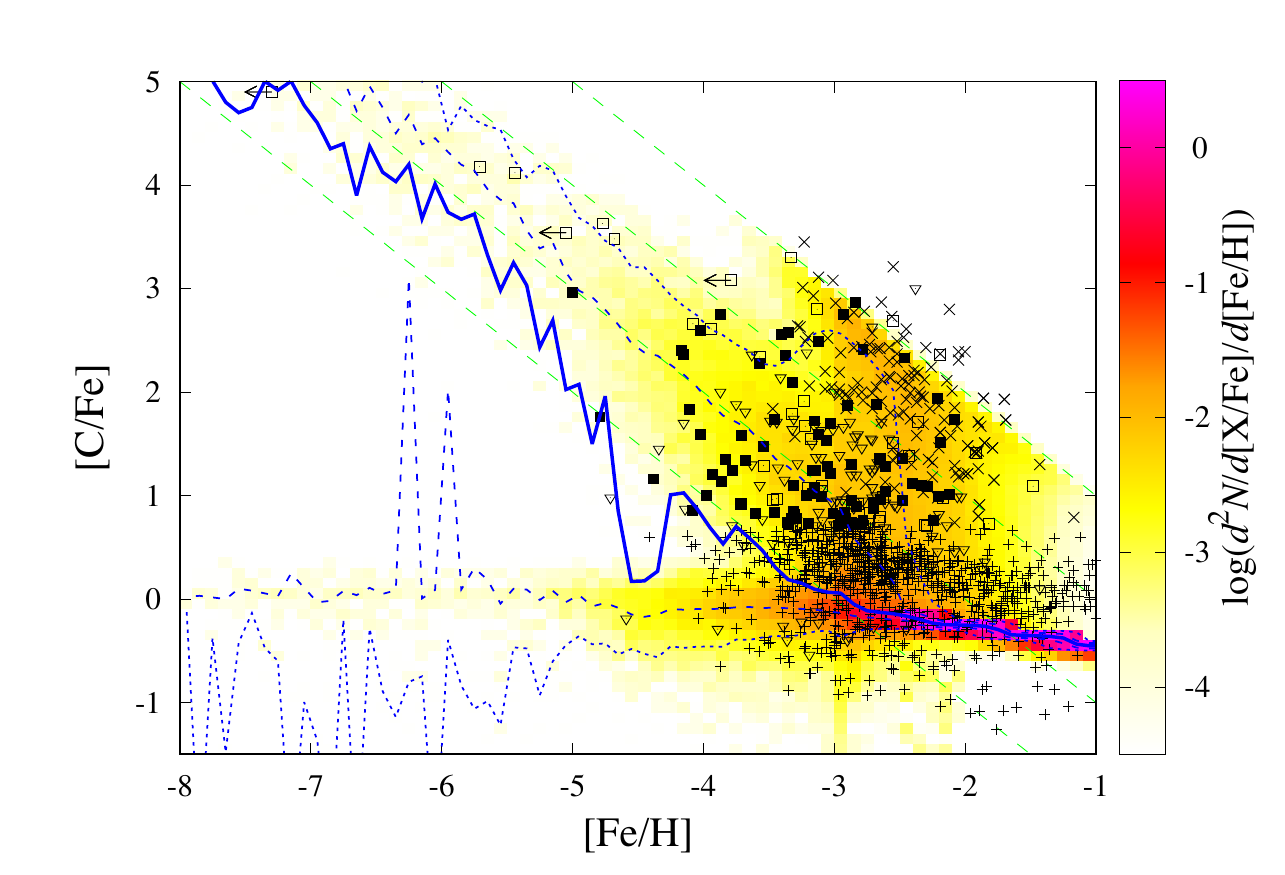}
\caption{
Distribution of carbon enhancement for the surface pollution by the binary mass transfer and the ISM accretion. We adopt the same period distribution of binaries as the Pop I stars for all the metallicity range (top panel) and the model with the period distribution of $P = 10^{4.5}$ days and $\sigma = 0.5$ at $\abra{Z}{H} < -3$ (bottom panel).
In these models, faint SNe are not considered. 
}\label{binary}
\end{figure}

We take into account the surface pollution by the accretion of ISM here, which is neglected in the previous sections.     
It has been considered as the main origin of iron on CRUMP stars in the binary scenario \citep{Suda04, Komiya15}. 
In this case, CRUMP stars are formed in a Pop III binary consisting of a low- or intermediate-mass primary star that can evolve to an AGB star, and a low-mass secondary that can survive to date in the nuclear burning stages. 
The binary mass transfer changes the surface carbon abundance of the secondary but does not affect its iron abundance, while the ISM accretion changes its surface iron abundance. 

Figure~\ref{binary} shows the results for the binary mass transfer and the ISM accretion scenario. 
In the top panel, we assume the same period distribution for all the metallicity range. 
The predicted abundance distribution is consistent with the observations for the metallicity of $\feoh \gtrsim -3$. 
As discussed in previous studies, many \cemps\ stars are formed up to $\coh \simeq 0$ through the binary mass transfer from AGB stars. 
Since the mass transfer also takes place at $\feoh \lesssim -3$, however, CEMP stars with $\coh \simeq 0$ are also formed at lower metallicity range. 

The Group I CEMP stars, not only \cemps\ stars but also CEMP-no stars with $\coh \gtrsim -1$, are very rare at $\feoh \lesssim -3.5$, which implies the lack of short period binaries that trigger the efficient mass transfer. 
In the bottom panel of Figure~\ref{binary}, we show the model with the optimum parameter set, $P = 10^{4.5} $ days and $\sigma_P = 0.5$, of the period distribution for the binaries in the low metallicity range of $\abra{Z}{H} < -3$. 
The majority of the HMP/UMP stars become CEMP stars with $-2 < \coh < -1$ by the binary mass transfer, which is consistent with the observed CRUMP stars.
It is to be noted that all the stars are assumed to be formed in binaries at $\abra{Z}{H} < -3$.
HMP stars without carbon enhancement are possible.
They are born with massive primary stars or belong to the binaries consisting of two low-mass stars that have not yet evolved to the AGB phase. 

If the surface iron pollution by ISM accretion is not taken into account, the predicted MDF is the same as the model without faint SN in Fig.~\ref{MDFfid} because the binary mass transfer does not change the iron abundance.  
As mentioned in \S~\ref{S:faintResult}, the predicted MDF is almost consistent with the observations except that a significant number of Pop III stars are expected to be observed instead of HMP/EMP stars. 
 
The surface pollution by ISM accretion change the surface metallicity of low-mass Pop III stars up to $\feoh \sim -5$ on average, which can be observed as HMP/UMP stars \citep[][]{Komiya15, Shen17}.    
CEMP-no stars in the Group~I are formed in short period binaries similar to \cemps\ stars but with the primary stars of $M \gtrsim 3.5 \msun$ since they have much smaller efficiency of $s$-process nucleosynthesis without the hydrogen ingestion by the helium flash convection as compared with \cemps\ stars with the primaries of $M < 3.5 \msun$ \citep{Suda17b,Yamada19a}.  

In summary, the binary mass transfer and the ISM accretion model with the change of binary period distribution at $\abra{Z}{H} = -3 \hyp -4$ well reproduces the distribution of carbon abundances in all the subgroups of CEMP stars. 
This scenario is also consistent with the observed MDF including the absence of Pop III stars.

%%% \S 5
\section{Conclusions}\label{S:conclusion}

We have investigated the origin of Carbon-Enhanced Metal-Poor stars with normal barium abundances (CEMP-no stars) using a chemical evolution model. 
We have updated the {\it StarTree} code, a chemical evolution model within the framework of the hierarchical galaxy formation, which is capable of tracing the inhomogeneous metal-enrichment process inside proto-galaxies. 
We consider two proposed scenarios for the origin of CEMP-no stars. 
One is the faint-supernovae (SN) scenario in which first supernova ejecta is rich in carbon relative to iron (or iron poor) to produce CEMP-no stars. 
The other is the binary scenario in which CEMP-no stars have accreted carbon-enhanced gas through the binary mass transfer from AGB companion stars. 

We find that the faint SN scenario for CEMP-no stars has severe difficulties in accounting for the abundance distribution of EMP stars. 
The predicted value of $\coh$ for the second generation stars is significantly lower than observed Carbon-Rich Ultra Metal-Poor (CRUMP) stars, where most of Hyper Metal-Poor (HMP) stars and Ultra Metal-Poor (UMP) stars are distributed below $\coh < -3$. 
The discrepancies will be alleviated by decreasing both the swept-up mass by SNe and the diffusion coefficient of SN yields by an order of magnitude.  
However, the model with inefficient metal mixing results in the overproduction of HMP/UMP stars. 
Adopting the high star-formation rate and the small mixing mass of SNe in combination with the high mass Initial Mass Function (IMF) of Pop III stars, we can reproduce the scarcity of carbon-normal stars observed at $\feoh \lesssim -4$.  
Another possibility to explain the high carbon abundance for HMP/UMP stars is the criterion for low-mass star formation depending on the carbon and iron abundances \citep{Chiaki18b}. 
In these two models, however, the averaged $\coh$ of HMP stars are lower than observed and the predicted number of UMP stars is significantly larger than observed.  

A fine tuning of the IMF at very low metallicity and extremely inefficient mixing of SN yields are demanded to explain the abundance distribution of HMP and UMP stars by the faint supernova scenario. 
Even under these assumptions, a mechanism other than faint supernova is required for CEMP-no stars with large carbon enhancement of $\coh > -1$ in the Group I defined by \citep{Yoon16}. 
It is only the Group II CEMP-no stars with weak carbon enhancement ($\cfe \sim +1$) that can be naturally reproduced by the faint SN models. 

In contrast, the binary mass transfer scenario is able to well reproduce the observed distribution of carbon and iron abundances only if we assume the change of the period distribution below $\abra{Z}{H} < -3$. 
This is demanded from the observed scarcity of \cemps\ stars at $\feoh < -3.5$. 

If CEMP-no stars are formed through the binary mass transfer, they provide the information on the formation and evolution of binaries in the extremely low-metallicity environment. 
Massive EMP or Pop III binaries are discussed as possible progenitors of binary black holes, observed by the gravitational waves \citep[e.g.][]{Kinugawa14}. 
They can also become high-mass X-ray binaries and play an important role for the cosmic reionization \citep[e.g.][]{Ricotti04, Jeon14}. 
Detailed studies on the evolution of binary mass transfers in CEMP-no stars are expected to reveal the nature of binary systems under the extremely low metallicity. 

\acknowledgments
We thank the anonymous referee for his/her useful comments and suggestions.
This work has been partially supported by a Grant-in-Aid for Scientific Research (JP23224004, JP15HP7004, JP19HP8019, JP16K05287, JP16K05298, JP16H02168, JP19K03931), from the Japan Society of the Promotion of Science.

\appendix

\section{Supernova Yields}

We test the SN yields by the ``piston models"of \citet{Heger10}.
In their models with some parameter sets, most of iron fall back onto the central compact object while carbon-rich layers in the outer shells are ejected. 
The predicted yields give large $\cfe$ as in the mixing and fallback model associated with faint SNe. 

They provide Pop III yields with a various set of progenitor mass, explosion energy, and mixing efficiency. 
We employ their S4 model with the mixing parameter set at zero.  
The explosion energy, $ E_{\rm SN} $, is determined randomly from their 10 models of $0.3 \hyp 10 \times 10^{51}$ erg. 

We adopt the yields for Pop III stars by \citet{Heger10} for $\abra{Z}{H} < -3$ and results of \citet{Woosley95} at higher metallicity.

%%% Fig. 18
\begin{figure}[htp]
\includegraphics[width=0.5 \columnwidth,pagebox=cropbox]{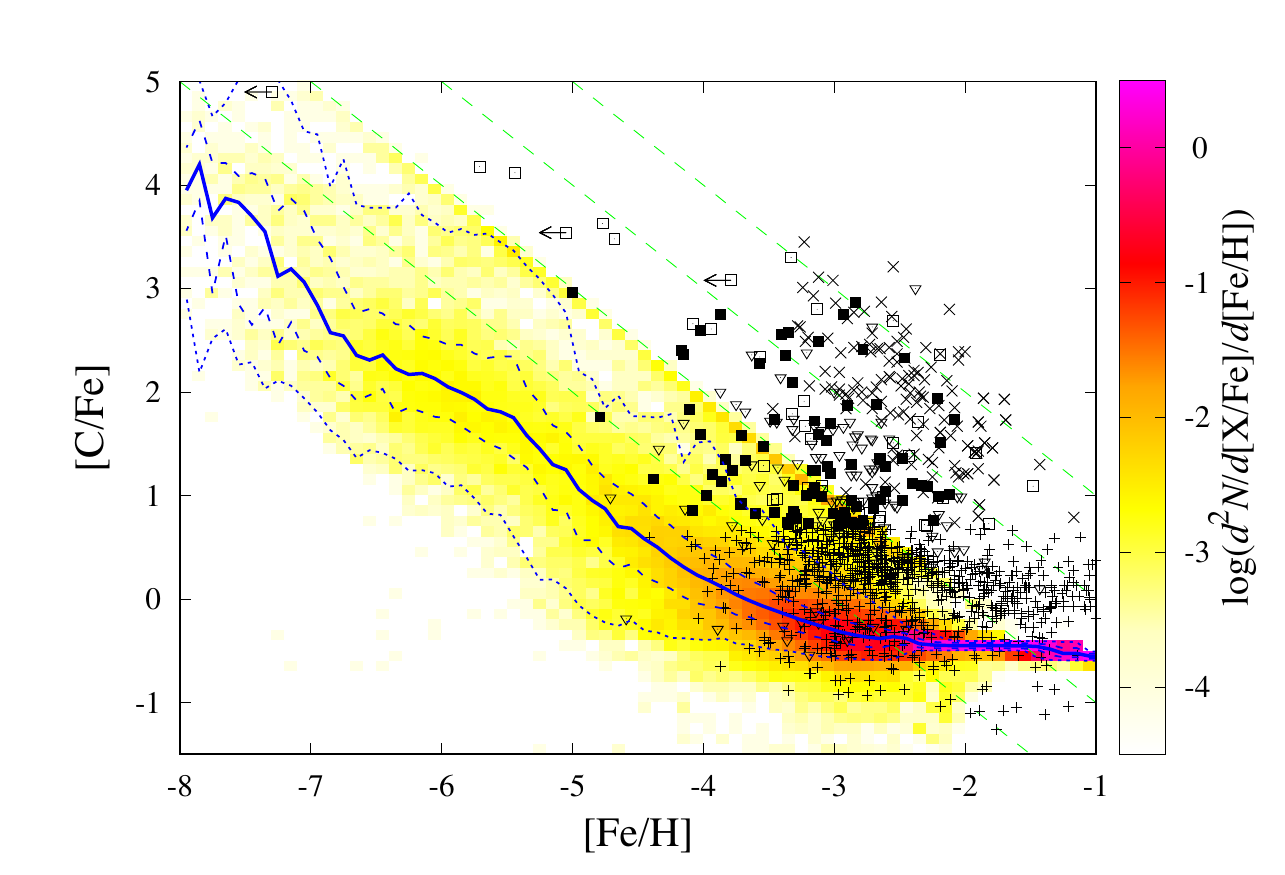} 
\includegraphics[width=0.5 \columnwidth,pagebox=cropbox]{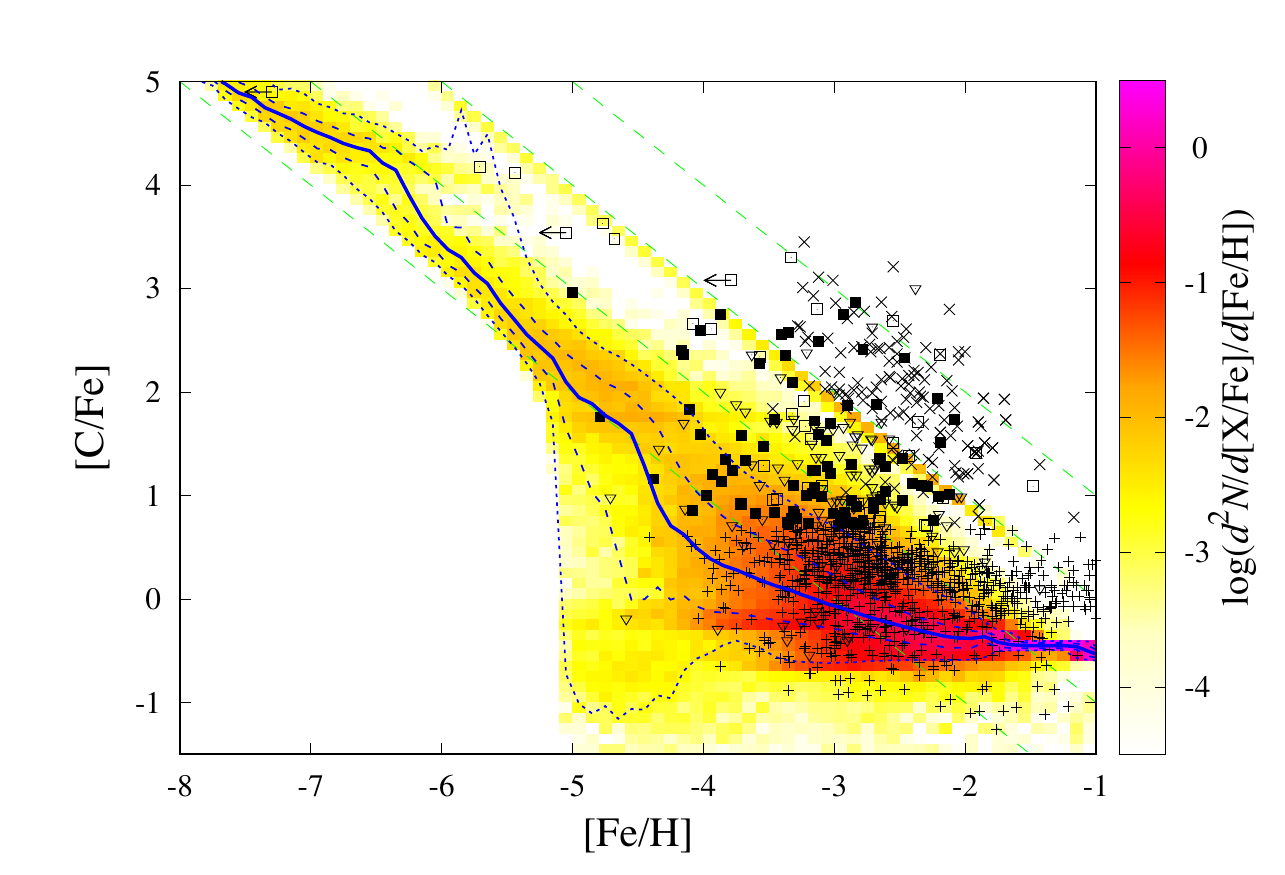} 
\caption{
Results with the SN yields by \citet{Heger10}. 
The SN explosion energy ($E_{\rm SN}$) is $0.3 \hyp 10 \times 10^{51}$ erg. 
Left panel: Result with the fiducial parameter set.  
Right panel: Result with $C_{\rm diff} = 10^{-7} \msun \, {\rm yr}^{-3/2}$, $ M_{\rm sw51} = 5.1 \times 10^{3} \msun $, and the low-mass star formation criterion by \citet{Chiaki18b}. 
}\label{Heger}
\end{figure}

%%% Fig. 19
\begin{figure}[htp]
\includegraphics[width=0.5 \columnwidth,pagebox=cropbox]{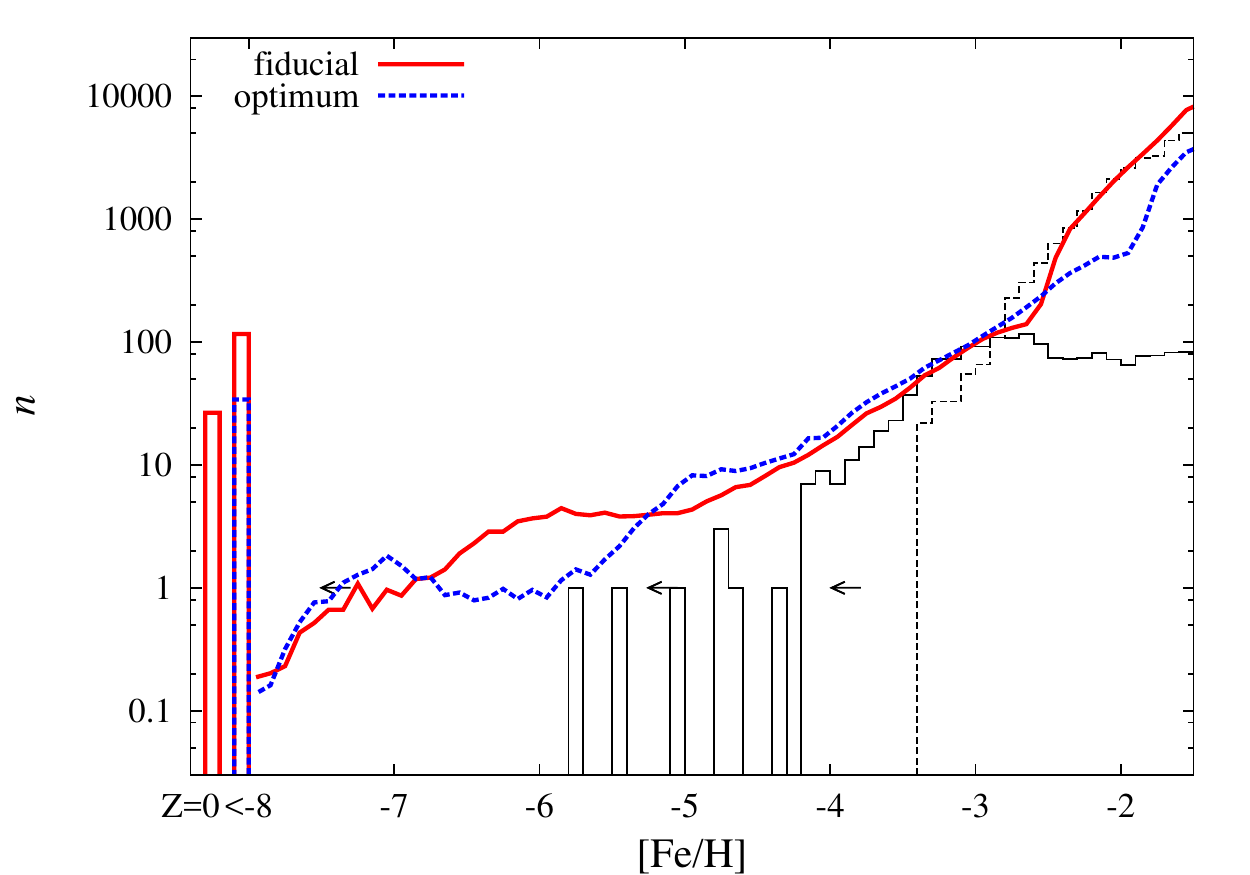} 
\caption{
MDFs of the model with the SN yields of \citet{Heger10}. 
The red and blue lines denote the models in the left and right panels of Figure~\ref{Heger}, respectively. 
}\label{MDF_Heger}
\end{figure}

%(fid)
Some stars with $m \gtrsim 40 \msun$ eject a large amount of carbon ($\Delta M_{\rm C} \gtrsim 1 \msun $) in the models of \citet{Heger10}. 
These SNe can produce the second generation stars with higher $\coh$ than the fiducial model in Section~\ref{S:result}. 
On the other hand, the majority of HMP/UMP stars also have $\coh < -3$ in the model with the fiducial parameter set, as shown in the left panel of Figure~\ref{Heger}. 

%(tuning)
We tried to reproduce the distribution of observed abundances by tuning the parameters.  
The right panel of Figure~\ref{Heger} is one of the best fit models. 
We adopt $C_{\rm diff} = 10^7 \msun \, {\rm yr}^{-3/2}$, $M_{\rm sw51} = 5.1 \times 10^3$, and the low-mass star formation criterion by \citet{Chiaki18b}. 
In this case, we find an inconsistency with the observations as in the optimum case with the mixing and fallback models in Section~\ref{S:optimum}. 
UMP stars are overproduced, and the Group I stars are not reproduced as discussed.

\end{document}